\documentclass[notitlepage,twocolumn,prx,nobalancelastpage,amssymb,superscriptaddress,showpacs,longbibliography,nofootinbib]{revtex4-1}

\usepackage{Symbols}
\usepackage{Shortcuts}
\usepackage{Text}
\usepackage[bottom]{footmisc}

\def \bk{{\boldsymbol{k}}}

\newcommand{\IE}[1]{\textcolor{black}{\textbf{}#1}}

\newcommand{\CY}[1]{\textcolor{black}{\textbf{}#1}}

\newcommand{\CYN}[1]{\textcolor{black}{\textbf{}#1}}

\newcommand{\CYNN}[1]{\textcolor{black}{\textbf{}#1}}

\newcommand{\CYL}[1]{\textcolor{black}{\textbf{}#1}}

\usepackage{stmaryrd}
\usepackage{epsfig}
\usepackage{subfigure}
\usepackage[colorlinks=true,citecolor=blue,linkcolor=blue]{hyperref}
\usepackage{verbatim}
\usepackage{xcolor}
\usepackage{siunitx}

\begin{document}
\title{Self-organized Floquet \IE{band geometry} 
in cavity-driven quantum materials}

\author{Christopher Yang}
\affiliation{Department of Physics and Astronomy, University of California, Irvine, California 92697, USA}
\affiliation{Institute of Quantum Information and Matter and Department of Physics,
California Institute of Technology, Pasadena, CA 91125, USA}

\author{Gil Refael}
\affiliation{Institute of Quantum Information and Matter and Department of Physics,
California Institute of Technology, Pasadena, CA 91125, USA}

\author{Mark S. Rudner}
\affiliation{Department of Physics, University of Washington, Seattle, WA 98195-1560, USA}

\author{Iliya Esin}
\affiliation{Department of Physics, Bar-Ilan University, 52900, Ramat Gan, Israel}

\date{\today}

\begin{abstract}
Floquet engineering has emerged as a powerful route to dynamically control band structure and topology in quantum materials, but most implementations rely on externally imposed laser fields that are power intensive, difficult to integrate into devices, and weakly coupled to the electronic system. 
\IE{We propose and analyze an alternative paradigm in which a self-generated cavity field Floquet-dresses the electronic bands and produces a geometric Hall response in an electrically driven cavity material system.} 
We consider a semiconductor layer embedded in a cavity and coupled to external leads and a bath of acoustic phonons, where dc pumping leads to the buildup of a coherent intracavity field through light-matter coupling. We determine the resulting nonequilibrium steady state self-consistently and show that, above threshold, the coupled system settles into a stable time-periodic limit cycle with a field amplitude set by the cavity quality factor and dissipation. \IE{This emergent periodic field Floquet-dresses the electronic bands and modifies the anomalous Hall response of a material with broken time-reversal symmetry.}
We demonstrate that the resulting Hall conductivity can be directly probed via in-plane dc transport measurements. \IE{Our work establishes a route to self organized Floquet band reconstruction and geometric transport without external laser illumination, highlighting cavity driven steady states as a platform for electrically controlled nonequilibrium phases.} 
\end{abstract}

\maketitle

\section{Introduction}

Controlling the electronic structure of quantum materials far from equilibrium has emerged as a powerful route to realizing phases of matter that are inaccessible in equilibrium \cite{Fausti2011Light-inducedCuprate,floquettopo,kitagawa,LiuTerahertzFieldInducedInsulator,doi:10.1126/science.1239834,Mankowsky2014,Mahmood2016,Mitrano2016,Flaschner2016ExperimentalBand,transport,PhysRevB.98.045104,Wong2018PullingField,fbe_adv,Oka2019,doi:10.1126/science.aaw4911,Topp2019TopologicalGraphene,Esin2020FloquetNanowires,Kumar2020LinearInsulators,VoglFloquetEngineeringOf2020,VoglEffectiveFloquet2020,VoglFloquetEngineeringOfTwistedDouble2020,Katz2020OpticallyGraphene,Chono2020Laser-inducedDichalcogenides,Li2020Floquet-engineeredGraphene,Nuske2020FloquetSolids,VoglFloquetEngineeringofTopological2021,Dehghani2021Light-inducedEngineering,Fazzini2021NonequilibriumLiquids,gyro,VoglFloquetEnginneering2021,Shan2021GiantEngineering,PhysRevB.105.174301,Castro2022FloquetTheory,KarniThroughMaterials,Vogl_2023,PhysRevLett.131.026901,PhysRevB.110.L100305,PhysRevLett.133.226301,Liu2025,Li2025,Choi2025,Persky2026,Wang2026}. In particular, periodic driving has enabled the dynamical engineering of band structures
\cite{doi:10.1126/science.1239834,Sentef2015FloquetBandFormation,Mahmood2016,Oka2019,floquethandbook,Rudner2020BandInsulators,KarniThroughMaterials,Choi2025,Wang2026},
symmetries and symmetry-broken phases
\cite{Frst2011,Mankowsky2014,PhysRevLett.117.090402,doi:10.1126/science.aaw4911,Harper2020,gyro},
and topology
\cite{Oka2009PhotovoltaicGraphene,PhysRevB.82.235114,floquettopo,PhysRevX.3.031005,Usaj2014IrradiatedInsulator,Rudner2020BandInsulators,PhysRevLett.131.026901},
giving rise to phenomena such as light-induced band inversions
\cite{doi:10.1126/science.1239834,Claassen2016AllOpticalTMD,Topp2019TopologicalGraphene,VoglFloquetEngineeringofTopological2021,PhysRevLett.131.026901,Wang2026},
Floquet topological insulators
\cite{PhysRevB.82.235114,floquettopo,PhysRevX.3.031005,PhysRevB.89.121401,Usaj2014IrradiatedInsulator,PhysRevLett.113.266801,Rudner2020BandInsulators},
and anomalous Hall responses in the absence of magnetic fields
\cite{Oka2009PhotovoltaicGraphene,kitagawa,PhysRevB.89.121401,Usaj2014IrradiatedInsulator,PhysRevLett.113.266801,Dehghani2015Out-of-equilibriumInsulator,Dehghani2016FloquetProduction,transport,Sato2019MicroscopicGraphene,Sato2019Light-inducedDissipation,McIver2020Light-inducedGraphene}. Within this Floquet paradigm, time-periodic modulation acts as an additional controlling knob, allowing the properties of solids to be reshaped on demand. 

To date, most realizations of Floquet engineering rely on externally imposed electromagnetic drives, typically provided by intense laser fields. While this approach has led to remarkable experimental progress, it also faces fundamental and practical challenges. Achieving appreciable Floquet gaps often requires large field amplitudes, which can induce unwanted heating, broadband excitations, and material damage \cite{Caruso2025,fbe_orig,fbe_adv,Aeschlimann2021SurvivalScattering,PhysRevB.79.184105}. \IE{Moreover, conventional Floquet schemes usually rely on intense optical fields supplied from outside the material. Delivering such fields locally and efficiently in compact or multi-element devices can be challenging.}

\begin{figure}[h!]
\includegraphics[width=1\columnwidth]{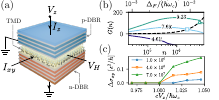}
\caption[]{\textbf{Setup and main results.}
(a) Schematic of the proposed cavity material device. A TMD monolayer is embedded between $p$- and $n$-doped distributed Bragg reflectors (DBRs), which both form an optical cavity resonant at frequency $\omega_c$ and serve as vertical electrical contacts biased by $V_z$. An in-plane probe current $I_{xy}$ is driven through the TMD, and the transverse Hall voltage $V_H$ is measured to extract the Hall conductivity $\sigma_{xy}$.
(b) Photon gain and loss as a function of the cavity photon occupation $n$, with the corresponding Floquet gap $\Delta_F$ indicated on the upper axis. Solid curves show the electronic gain $G(n)$ for several values of $eV_z$ slightly above $\hbar\omega_c$, while the dashed curve shows the cavity loss rate $\alpha n$. Rates are normalized by the lead tunneling rate $\Gamma^{\mathrm{lead}}$. The curve labels indicate $(eV_z/\hbar\omega_c-1)\times 10^3$. The self-consistent steady states are marked by the intersections of gain and loss.
(c) Photo-induced change in the Hall conductivity, $\Delta\sigma_{xy}$, as a function of $eV_z$ for several intrinsic cavity quality factors $Q$. A finite Hall response appears once the bias exceeds the photon energy, $eV_z>\hbar\omega_c$.}
\label{fig1}
\end{figure}

An alternative route to strong light-matter interaction is offered by optical and microwave cavities, which confine electromagnetic fields to small mode volumes and enhance their interaction with matter \cite{FriskKockum2019,RevModPhys.91.025005,Schlawin2022,RevModPhys.93.025005,PrezGonzlez2025}.
In cavity embedded materials, the relevant control parameter is no longer the free space field amplitude, but the cooperativity arising from the interplay of coupling strength, cavity lifetime, and electronic dissipation. This has enabled access to regimes where electronic states are strongly dressed by relatively modest photon populations, giving rise to polaritonic quasiparticles and cavity controlled material properties \cite{Zhou2024, Schneider2018,Santhosh2016,Basov2025,PhysRevResearch.2.033033,Dmytruk2022,PhysRevB.101.205140,PhysRevB.105.165121,Eckhardt2022,Schlawin2022,Hbener2020,Bloch2022}. 
Importantly, cavities also provide mode selectivity, which can stabilize coherent dynamics that would otherwise be washed out.

A closely related but conceptually distinct perspective comes from laser physics, where coherent electromagnetic fields 
emerge self-consistently from nonequilibrium conditions. In a laser, external pumping, gain, loss, cavity feedback, and nonlinear saturation combine to produce a stable limit cycle \IE{of the electromagnetic field,}
characterized by a well defined frequency and amplitude.
{From this perspective, we examine a new fundamental question}: can a 
dc-driven cavity-material system self-organize into a {time-}periodic steady state that 
supports \IE{Floquet modified Berry curvature and a geometric Hall response,}
without external laser illumination? Addressing this question requires going beyond conventional Floquet theory with externally fixed drives and instead treating the electromagnetic field and the electronic steady state on equal footing.

In this work, we develop such a framework. We consider an electrically driven material embedded in a cavity, where dc pumping leads to the buildup of a coherent intracavity field through light-matter coupling. \IE{We focus on a material with broken time reversal symmetry, which allows the self-consistent cavity field to acquire a definite chirality once the lasing instability sets in.} The electronic system is coupled to external leads and to phonons, providing excitation and dissipation channels that stabilize a nonequilibrium steady state. We determine this steady state self-consistently, accounting for the feedback between the electronic dynamics and the cavity field. Above threshold, the coupled system settles into a stable periodic steady state whose field amplitude saturates due to dissipation and is controlled by the cavity quality factor.
\IE{While we focus here on chirality generated by a time-reversal-breaking electronic medium, a related route could be realized by using a chiral or polarization-selective cavity, in which case the handedness of the periodic field is set by the photonic environment} \cite{Hbener2020,Voronin2022,PhysRevB.110.L121101,SurezForero2024,Tay2025,kulkarni2025realizationchiral}.

\IE{Crucially, this emergent chiral periodic field Floquet-dresses the electronic bands, reshaping their Berry curvature and nonequilibrium occupations,  }
\IE{and thereby modifying the transverse conductivity.}
Although the electronic distribution is nonthermal, the existence of a stationary periodic steady state allows transport coefficients to be defined and computed in a controlled manner. The Hall response can be directly probed via in-plane electrical contacts, providing a clear and experimentally accessible signature of cavity-induced Floquet \CY{band geometry.} 
In addition, optical and spectroscopic probes of the Floquet gap and Berry curvature offer complementary routes to detect \IE{the emergent Floquet dressed steady state.}

This approach offers several practical advantages over conventional externally driven Floquet schemes. Because the periodic field is generated self-consistently within the cavity, the required input power is set by the balance between gain and loss, rather than the  \CY{delivery of a high-intensity external drive}, 
\CY{providing an efficient route to Floquet engineering}. 
Moreover, the use of dc electrical pumping provides a natural route toward device integration and scalability, allowing for compact, on-chip implementations of Floquet-engineered phases without the need for high-power optical infrastructure.

\CYNN{Our analysis proceeds in four steps. In Sec.~\ref{sec:physsetup}, we introduce the physical setup and formulate the effective cavity dynamics, including the nonlinear gain function \(G(n)\) that determines the self-consistent photon population. In Sec.~\ref{sec:elpopdyn}, we derive \(G(n)\) within a phenomenological electronic population model, which exposes the mechanisms of gain, saturation, and the approach to a Floquet-dressed distribution with an enhanced anomalous Hall response. In Sec.~\ref{sec:micmodel}, we develop a microscopic Floquet--Boltzmann transport model for a magnetically doped TMD and use it to compute the steady-state electronic distribution, photon gain, and self-consistent Floquet gap. Finally, in Sec.~\ref{sec:signaturesss}, we discuss experimental signatures of the self-organized Floquet steady state, including the anomalous Hall response, the out-of-plane current response to bias quenches, optical probes of the Floquet gap, and possible edge-state transport signatures.}

\section{Physical Setup and Steady State Feedback} \label{sec:physsetup}

We consider a two-dimensional semiconductor embedded inside a single-mode electromagnetic cavity, as schematically illustrated in Fig.~\ref{fig1}(a). The active material is a transition-metal dichalcogenide (TMD) monolayer placed inside a vertical optical resonator formed by electrically contacted Bragg reflectors, in an architecture analogous to that of a vertical cavity surface emitting laser (VCSEL) \cite{Jewell1989,902168}. The same vertical contacts that inject current into the \IE{TMD layer} 
also act as mirrors for the cavity mode, confining the electromagnetic field via Bragg reflection.
In addition, independent horizontal contacts are attached to the two-dimensional layer and are used to probe in-plane transport properties of the nonequilibrium steady state, including the Hall conductivity associated with the self-organized Floquet state.

\IE{
We focus on a TMD in which time reversal symmetry is weakly broken, for example by dilute magnetic impurities. This lifts the equivalence between the two valleys, so that under suitable pumping conditions the optical response is dominated by a single active valley. Since the two valleys of a TMD couple selectively to opposite circular polarizations, the electronic gain becomes helicity dependent. As a result, above threshold, one helicity can develop a coherent amplitude, while the opposite helicity remains below threshold.}

\IE{After this polarization selection, the active electromagnetic degree of freedom is effectively described by a single chiral cavity mode. We model this mode as
$\hat H_{\mathrm{cav}} = \hbar \omega^{(0)}_c \hat a^\dagger \hat a$,
where $\omega^{(0)}_c$ is the bare cavity frequency and $\hat a^\dagger$, $\hat a$ are photon creation and annihilation operators for the selected circular polarization. We therefore restrict our analysis to this chiral mode and neglect the opposite polarization in the following.}

\CYNN{In the remainder of this section, we formulate the general cavity dynamics by parametrizing the net photon production of the electronic medium through a
gain function \(G(n)\), without yet specifying its microscopic origin. We then specify how a finite coherent cavity field, once present, acts back on the
electronic system: it Floquet-dresses the electronic bands, opens a resonant Floquet gap, and modifies the Berry curvature and Hall response. The physical
mechanisms that generate photon gain, saturate the cavity field, and determine \(G(n)\) self-consistently from the electronic steady state are then developed in
Secs.~\ref{sec:elpopdyn} and~\ref{sec:micmodel}.}

\subsection{Cavity photon population dynamics}

\IE{We assume that, above threshold, the selected chiral cavity mode develops a coherent classical field \cite{Haken1963,PhysRev.134.A1429,PhysRev.159.208,Graham1968}}. Its expectation value can be written as
\begin{equation}
\langle \hat a(t)\rangle = A e^{-i\omega_c t},
\label{eq:cavity_coherent_field}
\end{equation}
where \(A\) is a slowly varying complex amplitude and \(\omega_c\approx\omega^{(0)}_c\) is close to the bare cavity resonance frequency. In the steady state considered below, the average photon occupation associated with this coherent field is $n =|A|^2$. 

\CYNN{Our analysis is restricted to the semiclassical regime in which the steady-state photon number is large, \(n\gg1\). In this regime, relative amplitude fluctuations and phase diffusion are parametrically suppressed by factors of \(n^{-1/2}\) and $n^{-1}$  \cite{Henry1982TheoryOT,PhysRev.112.1940}, respectively.\footnote{\CYNN{The phase-diffusion rate is given by $D_\phi \sim \frac{\omega_c}{2Qn}(1+\alpha_H^2)$, where $Q$ is the quality factor of the cavity and \(\alpha_H\) is the Henry factor \cite{Henry1982TheoryOT,PhysRev.112.1940}. For representative values \(n\gtrsim10^5\), \(Q\gtrsim10^6\), \(\hbar\omega_c\sim1\,{\rm eV}\), and \(\alpha_H\sim5\), \(D_\phi^{-1}\) is in the $\mu \mathrm{s}$ range or longer, far exceeding the typical ps-scale electronic relaxation times \cite{fbe_orig}. Therefore, on the timescales relevant for the electronic steady state, the cavity field can be treated as a coherent classical oscillation, as in Eq.~\eqref{eq:cavity_coherent_field}.}}}
\CYNN{Under this condition}, the photon occupation evolves according to the rate equation \cite{Haken1963,PhysRev.134.A1429,PhysRev.159.208,Graham1968,PhysRev.145.110,Haken1966,Graham1968,Henry1982TheoryOT}
\begin{equation}
\dot n = G(n)-\alpha n ,
\label{eq:rate_eq}
\end{equation}
where \(\alpha\) denotes the cavity loss rate, related to the cavity quality factor by \(Q=\omega_c/\alpha\). 

The gain term \(G(n)\) accounts for electronic transitions induced by the dc-driven nonequilibrium steady state, which can either emit photons into the cavity mode through stimulated emission or absorb photons from the cavity. \IE{Positive \(G(n)\) corresponds to net gain, while negative \(G(n)\) corresponds to net absorption.} Crucially, \(G(n)\) depends on the electronic state of the material, which itself is influenced by the coherent cavity field. We further assume a separation of timescales in which the electronic occupations relax to their nonequilibrium steady state much faster than the cavity photon number changes.\footnote{\CYNN{The lifetime of the cavity field can be approximated by \(\tau_{\rm cav}=Q/\omega_c\sim0.7\!-\!70\,{\rm ns}\)
for \(\hbar\omega_c\sim1\,{\rm eV}\) and \(Q=10^6\!-\!10^8\). This is much longer than the typical ps-scale
electronic relaxation times \cite{fbe_orig}.}}
This scale separation gives \CYNN{an effective} gain function \(G(n)\) that is generally a nonlinear function of \(n\), see Fig.~\ref{fig1}(b).

\IE{
We use this nonlinear gain function to identify the self consistent operating point of the coupled cavity material system, by setting $\dot n=0$ in Eq.~\eqref{eq:rate_eq}. This fixes the coherent field amplitude, and hence the electric field amplitude \CYNN{is given by} \begin{equation}
\mathcal{E} = \mathcal{E}_0 |A|, \qquad \mathcal{E}_0 = \sqrt{\frac{\hbar\omega_c}{2 \epsilon \epsilon_0 \mathcal{V}}},
\label{eq:field_strength}
\end{equation}
where $\epsilon_0$ is the vacuum permittivity, $\epsilon$ is the relative permittivity, and $\mathcal{V}$ is the effective cavity mode volume. \CYNN{For the cavity to reach a finite field amplitude, the electronic medium must provide positive gain over a range of photon occupations such that \(G(n)\) exceeds the cavity loss rate \(\alpha n\) before saturating [see Eq.~(\ref{eq:rate_eq})]. This gain is enabled by applying a voltage bias \(V_z\) across the vertical electrical contacts [see Fig.~\ref{fig1}(a)]. In particular, when the bias energy exceeds the cavity photon energy, \(eV_z>\hbar\omega_c\), the filtered contacts generate a population inversion near the photon resonance, so that stimulated emission into the selected cavity mode can dominate absorption.} 
}

\subsection{\CYNN{Effective electronic Floquet problem}}

Once a finite cavity field is established, the field acts back on the electronic system as an emergent periodic drive, Floquet-dressing the bands, reshaping their Berry curvature, and modifying the Hall transport response. We now describe these effects in detail.

In the semiclassical regime, where Eq.~\eqref{eq:cavity_coherent_field} is valid, the electronic dynamics \CYN{in the cavity} is governed by the effective time-periodic Hamiltonian, 
\begin{equation}
\hat H_{\mathrm{el}}(t) = \hat H_{0,\mathrm{el}} +
 \hat{O}_{e-p} A e^{-i\omega_c t}+h.c.
\label{eq:effective_floquet_ham}
\end{equation}
Here, $\hat H_{0,\mathrm{el}}$ describes the equilibrium band structure of the semiconductor and $\hat{O}_{e-p}$ is the electron--cavity--photon coupling operator, whose form depends on the polarization of the cavity mode. 

For a translationally invariant system, Eq.~\eqref{eq:effective_floquet_ham} can be written in momentum space as
$
\hat H_{\mathrm{el}}(t)
=
\int \frac{d^2 k}{(2\pi)^2}
\,
\hat{\boldsymbol{\psi}}_{\mathbf{k}}^\dagger
H_{\mathrm{el}}(\mathbf{k},t)
\hat{\boldsymbol{\psi}}_{\mathbf{k}}$,
where \(H_{\mathrm{el}}(\mathbf{k},t)\) is the corresponding single-particle Bloch Hamiltonian and \(\hat{\boldsymbol{\psi}}_{\mathbf{k}}\) is a spinor of electronic annihilation operators in the relevant band, spin, and valley basis.

\begin{figure}[t]
\includegraphics[width=1\columnwidth]{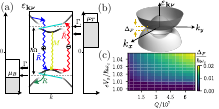}
\caption[]{\textbf{Induced Floquet spectrum.}
(a) Floquet quasienergy spectrum and scattering processes responsible for gain saturation. The biased top and bottom leads, with chemical potentials $\mu_T$ and $\mu_B$, inject and remove carriers near the photon resonance $\varepsilon_{\boldsymbol{k}\nu}=\pm\hbar\omega_c/2$. Recombination, with rate $M$, and phonon relaxation, with rate $R$, compete with photon-assisted Floquet-Umklapp processes that change the cavity photon number by $\pm 1$ and occur with rate $\tilde R=R(\Delta_F/\hbar\omega_c)^2$. 
(b) Floquet band structure showing the cavity-induced gap $\Delta_F$.
(c) Steady-state gap $\Delta_F$ as a function of the cavity quality factor $Q$ and bias $eV_z=\mu_T-\mu_B$.}
\label{fig2}
\end{figure}

The Hamiltonian \(H_{\mathrm{el}}(\mathbf{k},t)\) is periodic in time with period \(T=2\pi/\omega_c\). The electronic states can therefore be expressed in terms of Floquet eigenstates \(|\phi_{\mathbf{k}\nu}(t)\rangle\), satisfying
\begin{equation}
\left[
H_{\mathrm{el}}(\mathbf{k},t)
-
i\hbar\partial_t
\right]
|\phi_{\mathbf{k}\nu}(t)\rangle
=
\varepsilon_{\mathbf{k}\nu}
|\phi_{\mathbf{k}\nu}(t)\rangle ,
\end{equation}
where \(\varepsilon_{\mathbf{k}\nu}\) is the quasienergy of the Floquet state labeled by momentum \(\mathbf{k}\) and band index \(\nu\). The Floquet eigenstates are periodic in time:
\[
|\phi_{\mathbf{k}\nu}(t+T)\rangle
=
|\phi_{\mathbf{k}\nu}(t)\rangle ,
\]
\CYNN{and the quasienergies $\varepsilon_{\boldsymbol{k}\nu}$ are defined modulo \(\hbar\omega_c\).}

Physically, the coherent cavity field hybridizes electronic states \CYNN{most strongly when their energetic separation is close to} integer multiples of \(\hbar\omega_c\). In the regime where the cavity frequency is close to the direct interband transition, the dominant effect is the resonant hybridization of the valence band with the one-photon replica of the conduction band \CYL{[see Fig.~\ref{fig2}(a)]}. Below \CYNN{the voltage bias threshold, $eV_z \leq \hbar\omega_c$}, \CYNN{the cavity amplitude vanishes} (\(A=0\)), the Floquet replicas decouple and the equilibrium band structure is recovered. Above threshold \CYNN{($eV_z > \hbar\omega_c$)}, the self-organized cavity field opens a Floquet gap \CY{$\Delta_F\propto |A|$} \CY{\CYL{in the single-particle spectrum} along the resonance ring, i.e., the closed curve in 
momentum space where the valence band is resonant with the one-photon 
replica of the conduction band.}
\CY{This Floquet description is valid in the limit}
\begin{equation} \label{eq:flcond}
\hbar / (\tau_{\mathrm{rel}} \, \Delta_F) \ll 1 ,
\end{equation}
\CY{where $1/\tau_{\mathrm{rel}}$ denotes the total scattering rate of Floquet states,} \IE{which takes into account} 
\CY{the electronic scattering rate,} \IE{the cavity quality factor, \CYNN{and the phase diffusion time of the cavity field}.} 
\CYNN{Note that the relaxation time $\tau_{\rm rel}$ is suppressed by Pauli blocking in a Floquet system, and is therefore much slower than the bare scattering rate of electrons \cite{fbe_orig}.\footnote{For a representative
electronic relaxation time \(\tau_{\rm rel}=10\,{\rm ps}\) and near-infrared photon energy \(\hbar\omega_c=1\,{\rm eV}\), Eq.~(\ref{eq:flcond}) gives $\Delta_F / (\hbar\omega_c) \gg 10^{-4}$.  Our numerics [see, e.g., Fig.~\ref{fig1}(b)] focus on this regime where the Floquet gap, Berry-curvature and transport response are well-resolved.}}

\subsection{Geometric signatures of the Floquet states} \label{sec:geomsig}

The Floquet spectrum determines both the electronic transition rates and the geometric response of the steady state \CYL{together with the steady-state electronic populations}. In Fig.~\ref{fig2}(a), we show \CYNN{an example of} the period-averaged spectral function of \(H_{\mathrm{el}}(\mathbf{k},t)\), which exhibits the Floquet gap near the resonance. \CY{Here, we have focused on an effective two-band system near one of the optically-active valleys of the TMD, which can be described by a massive Dirac Hamiltonian with a gap slightly below $\hbar\omega_c$.}

The folded quasiband structure, sketched in Fig.~\ref{fig2}(b), can acquire a drive-induced Berry curvature and, depending on the occupation of the Floquet bands, an \CYL{enhanced} transverse Hall response. In particular, each Floquet band $\nu$ is characterized by a Berry curvature
\begin{equation}
\mathcal{B}_{\mathbf{k}\nu}{(t)}
=
i \left(
\langle \partial_{k_x} \phi_{\mathbf{k}\nu} |
\partial_{k_y} \phi_{\mathbf{k}\nu} \rangle
-
\langle \partial_{k_y} \phi_{\mathbf{k}\nu} |
\partial_{k_x} \phi_{\mathbf{k}\nu} \rangle
\right),
\end{equation}
defined in terms of the periodic part of the Floquet eigenstate
$|\phi_{\mathbf{k}\nu}(t)\rangle$.
The Floquet hybridization induced by the cavity field strongly modifies $\mathcal{B}_{\mathbf{k}\nu}{(t)}$ {relative to the original Berry curvature of the band structure in the absence of driving}, concentrating Berry curvature near the Floquet resonance ring where the gap $\Delta_F$ opens \cite{floquettopo,transport}, see Fig.~\ref{fig4}(b).

\IE{The period-averaged Berry curvature of the Floquet quasibands, \CYN{$\overline{\mathcal B}_{\mathbf k\nu} = \frac{1}{T}\int_0^T dt\, \mathcal B_{\mathbf k\nu}(t)$}, can be used to define a band Chern number,
\begin{equation}
C_{\nu}
= 
\frac{1}{2\pi} \int d^2 \mathbf{k} \ 
\overline{\mathcal B}_{\mathbf k\nu}.
\end{equation}
\CYNN{The Chern number measures the (oriented) net number of chiral edge modes crossing the quasienergy gap above and below the Floquet band \cite{PhysRevX.3.031005}.}
For the corresponding fully driven two-valley model, including the relevant spin and valley sectors, the Floquet bands will carry a total Chern number.  Therefore, in the limit where the Floquet gap is well resolved 
this quasiband topology is expected to support chiral edge modes, sketched in Fig.~\ref{fig2}(a).}

\IE{A central consequence of the drive-induced Berry curvature is an in-plane anomalous Hall conductivity. In the nonequilibrium steady state, this response is obtained by integrating the period-averaged Berry curvature weighted by the Floquet state occupations ($f_{\mathbf{k}\nu}$),
}
\begin{equation}
\sigma_{xy}
= 
\frac{e^2}{\hbar}
\sum_{\nu}
\int \frac{d^2 \mathbf{k}}{(2\pi)^2}
f_{\mathbf{k}\nu}\,
\overline{\mathcal B}_{\mathbf k\nu}.
\end{equation}
\CYL{The steady-state Floquet occupations are solved self-consistently with the cavity rate equation, Eq.~(\ref{eq:rate_eq}), whose gain function $G(n)$ is implicitly a function of $f_{\mathbf{k} \nu}$. The relation between $G(n)$ and $f_{\mathbf{k} \nu}$, in particular, is obtained using the Floquet--Boltzmann transport equation, which we motivate phenomenologically in Sec.~\ref{sec:elpopdyn} and solve numerically in Sec.~\ref{sec:micmodel}.}

In practice, we focus on the photoinduced change $\Delta\sigma_{xy}$ relative to the equilibrium value, which isolates the contribution arising from Floquet band dressing.
Fig.~\ref{fig1}(c) shows $\Delta\sigma_{xy}$ as a function of bias voltage \CYNN{$V_z$} for several cavity quality factors\CYL{. Here, the Floquet occupations used to compute $\Delta\sigma_{xy}$, were obtained from the microscopic Floquet-Boltzmann equation detailed in Sec.~\ref{sec:micmodel}}. The Hall response remains small below threshold $e V_z\simeq \hbar\omega_c$ before increasing rapidly in the regime $e V_z> \hbar\omega_c$, where a finite self-organized Floquet gap develops. Its magnitude also increases with the cavity quality factor, corresponding to an enhancement of the Floquet gap. Our goal in the following sections is to characterize this sharp modification of the Hall conductivity and explain the origin of gain, saturation, and the self-consistent Floquet gap.

\section{Electronic population dynamics and gain function} \label{sec:elpopdyn}
We now explain how the gain function, \(G(n)\), is obtained from the electronic steady state. Since the photon number evolves slowly compared with the electronic relaxation dynamics, we treat \(n\) as a quasi-static parameter. For each value of \(n\), we compute the nonequilibrium steady state of the electronic system and evaluate the resulting net rate of photon emission into, and absorption from, the selected cavity mode. This procedure defines the nonlinear gain function \(G(n)\) that appears in Eq.~\eqref{eq:rate_eq}.

\IE{The gain is generated by a vertical dc bias applied through the doped cavity mirrors. The \(n\)- and \(p\)-doped contacts respectively inject electrons into conduction-band states and extract electrons from valence-band states, thereby creating a nonequilibrium population of electrons and holes in the TMD. This selectivity is modeled by contacts whose densities of states are confined to finite, nonoverlapping energy windows, indicated by the black boxes in Fig.~\ref{fig2}(a). In practice, such energy filtering can be realized using semiconducting contact materials with \CYNN{narrow band widths and} band gaps much larger than the cavity photon energy \cite{fbe_orig,transport,PhysRevB.110.075428,PhysRevB.96.165443}. The voltage bias, \(eV_z=\mu_T-\mu_B\), is controlled by the chemical potentials of the top and bottom leads. For \(eV_z>\hbar\omega_c\), this biased electronic steady state can provide net stimulated emission into the selected chiral cavity mode, and a finite photon occupation develops once the gain exceeds cavity loss.}

\subsection{Phenomenological model for the gain function} \label{sec:phenomgain}

\begin{figure}[t] \includegraphics[width=1\columnwidth]{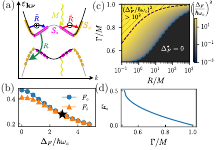} \caption[]{\textbf{Phenomenological model.} (a) Folded Floquet-band representation of the scattering processes shown in the extended-zone picture in Fig.~\ref{fig2}(a). The patches $S_o$ and $S_i$ denote states near the resonance ring in the upper Floquet band, while $S_{o'}$ and $S_{i'}$ are their particle-hole counterparts in the lower band. Solid arrows indicate photon-assisted phonon processes that emit $(+)$ or absorb $(-)$ a cavity photon, with rate $\tilde R=R(\Delta_F/\hbar\omega_c)^2$, and squiggly arrows indicate recombination processes with rate $M$. (b) Average occupations $F_o$ and $F_i$ as a function of the Floquet gap $\Delta_F$. (c) Steady-state gap $\Delta_F^*$ as a function of phonon scattering rate $R$ and lead tunneling rate $\Gamma$, both normalized by the recombination rate $M$. The dark region indicates where no finite-gap solution exists. (d) Steady-state occupation $F_i=F_o\equiv F$ as a function of $\Gamma/M$.} \label{fig3} 
\end{figure}

\CYN{We first present a minimal phenomenological model whose purpose is to expose the mechanism of gain, saturation, and emergence of an anomalous Hall conductivity. The full microscopic Floquet--Boltzmann calculation is presented in Sec.~\ref{sec:micmodel}.}
The phenomenological model focuses on \CY{the optically active bands near one of the valleys of the TMD, which we assume to be particle-hole symmetric}. \CY{The model}
focuses on the electronic states near the Floquet resonance, which are shown schematically in Fig.~\ref{fig3}(a). 
For convenience we define an inner patch \(S_i\), to denote states in the upper Floquet band inside the resonance ring, and an outer patch \(S_o\), to denote states in the upper Floquet band outside the resonance ring. \CY{The inner and outer patches in the lower Floquet band are denoted \(S_{i'}\) and \(S_{o'}\), respectively.} 
The phenomenological variables are the \CY{band-resolved average} occupations of the upper Floquet band in the two patches, $F_i$, $F_o$.
\CY{We \CYNN{assume} charge neutrality, \CYNN{particle-hole symmetric bands, and particle-hole symmetric scattering processes,} where the occupations in the lower Floquet band are particle-hole symmetric},
$F_{i'}=1-F_i$, $F_{o'}=1-F_o$.

The non-equilibrium electronic steady state 
arises from a balance between energy injection and dissipation, {together with the coherent modification of the system's (Floquet) bands in the regime where a strong field builds up in the cavity}. Energy is pumped into the electronic system by coupling to \IE{energy-filtered vertical} leads under an applied dc bias, while energy is dissipated through phonon emission and through photon loss from the cavity\IE{, as illustrated in Fig.~\ref{fig2}(a)}. 
The key scattering processes governing the patch occupations are indicated in Fig.~\ref{fig2}(a) in the extended zone picture and in  Fig.~\ref{fig3}(a) in the band-folded picture. \CYNN{The two} pictures are equivalent, yet provide different perspectives on the problem.

\CYL{The source lead, with chemical potential $\mu_T$, injects electrons into states derived from the conduction band of the system in the absence of the coherent field. In the band-folded picture, this means that the lead injects electrons into the outer patch $S_o$, with rate $\dot F_o|_{\mathrm{lead}}=\Gamma (1-F_o)$. Similarly, the drain lead, with chemical potential $\mu_B$, extracts electrons from states near the top of the original valence band, and hence from the inner patch $S_i$. We assign a rate $\dot F_i|_{\mathrm{lead}}=-\Gamma F_i$ to this process.} Here $\Gamma$ denotes the average tunneling rate, and is an increasing function of \IE{the bias voltage,} $V_z$.

Focusing on \CYL{further} processes that enter the rate equation for \(F_i\), electrons in \(S_i\) can relax through several channels. First, an electron in \(S_i\) can scatter to the lower-band patch \(S_{o'}\) through ordinary phonon emission [shown in Figs.~\ref{fig2}(a) and \ref{fig3}(a)]. This process does not exchange photons with the cavity mode and contributes
$\dot F_i|_{\rm ph}
=
-R F_i F_o$, 
where the factor \(F_o\) follows from the availability of holes in \(S_{o'}\), using \(F_{o'}=1-F_o\), and \(R\) is the average electron-phonon scattering rate.

Second, \(F_i\) is affected by photon-assisted Floquet--Umklapp processes, which exchange a single photon with the selected cavity mode. In the folded-band picture, these processes correspond to transitions between Floquet sectors whose indices differ by one. In the physical, extended-zone picture, they correspond to scattering processes in which a cavity photon is either emitted or absorbed, accompanied by phonon emission or another relaxation process that provides the remaining energy and momentum transfer. Since these transitions are enabled by the coherent mixing of Floquet sectors, their rates are suppressed by the factor \((\Delta_F/\hbar\omega_c)^2\).

The photon-absorption processes that deplete \(S_i\) are \(S_i\to S_o\) [shown in Figs.~\ref{fig2}(a) and \ref{fig3}(a)] and \(S_i\to S_{i'}\) (see SI). Their combined contribution to the occupation of \(S_i\) is
$\dot F_i|_{\rm abs}
=
-R\left(\frac{\Delta_F}{\hbar\omega_c}\right)^2
\left[
F_i(1-F_o)+F_i^2
\right]$.
The reverse photon-emission processes populate \(S_i\) with the rate
$\dot F_i|_{\rm em}
=
R\left(\frac{\Delta_F}{\hbar\omega_c}\right)^2
F_o(1-F_i)$, corresponding to scattering from \(S_o\) to \(S_i\). 

Finally, radiative recombination into non-cavity modes and Auger-like processes are modeled phenomenologically by an interband relaxation rate \(M\). \CYNN{In the unfolded-band picture [Fig.~\ref{fig2}(a)] recombination corresponds to the usual relaxation of an electron-hole pair across the semiconductor gap. In the folded-band picture [Fig.~\ref{fig3}(a)], the same process appears as a transfer from the lower-band partner patch \(S_i'\) into the upper-band patch \(S_i\), giving} 
$\dot F_i|_{\rm inter}
=
M(1-F_i)^2$.
This process conserves the cavity photon number and therefore does not contribute directly to the gain function, but it affects the electronic steady state that determines the gain.
Collecting the processes discussed above, the rate equation for \(F_i\) is written as
\begin{equation}
\label{eq:ssfi}
\dot F_i
=
\dot F_i|_{\rm lead}
+
\dot F_i|_{\rm ph}
+
\dot F_i|_{\rm abs}
+
\dot F_i|_{\rm em}
+
\dot F_i|_{\rm inter}.
\end{equation}

The rate equation for \(F_o\) is obtained in the same way, by summing all processes that populate or deplete the outer patch \(S_o\). The lead contribution injects electrons into \(S_o\), while the ordinary phonon-assisted terms are the counterparts of those appearing in \(\dot F_i\). Similarly, the photon-assisted Floquet--Umklapp terms include both absorption and emission processes, which we denote by \(\dot F_o|_{\rm abs}\) and \(\dot F_o|_{\rm em}\), respectively. These terms are obtained from the same scattering channels described above, with the appropriate signs according to whether they populate or deplete \(S_o\).

Finally, the gain function for the cavity photons is obtained from the photon-assisted Floquet--Umklapp processes. These processes either emit a photon into the selected cavity mode or absorb a photon from it. We define the total photon-emission and photon-absorption rates as $\tilde R_{\rm em}=  - \dot F_o\big|_{\rm em}$, $\tilde R_{\rm abs} = - \dot F_i\big|_{\rm abs}$, respectively. The net photon gain is therefore given by
\begin{equation}
G(n) = \tilde R_{\rm em} - \tilde R_{\rm abs}. 
 \label{eq:phenom_gain_def}
\end{equation}

\subsection{Steady State Solution}
In a perfect cavity with no photon loss, the steady-state condition
\(G(n)=0\) corresponds to a balance between photon-emission and
photon-absorption processes, see Figs.~\ref{fig2}(a) and
\ref{fig3}(a). In the phenomenological patch model, this balance is
possible only when the inner and outer patches have equal occupation,
\(F_i=F_o\). The conditions for such a steady state can be obtained by
solving the electronic rate equation \(\dot F_i=0\) [Eq.~(\ref{eq:ssfi})]
under the constraint \(F_i=F_o\), which gives
\begin{equation} \label{eq:foficond}
F_i = F_o \equiv F = \frac{1-\sqrt{2\Gamma/M - 1}}{2}.
\end{equation}
Here the occupation is controlled by the ratio of the electronic pumping
rate \CYL{from the leads,} \(\Gamma\), to the recombination rate \(M\). For this solution to be
physical, the occupation must satisfy \(0\leq F\leq 1\), which restricts
the relevant parameter regime to
\begin{equation}
\frac{1}{2}<\frac{\Gamma}{M}\leq 1 .
\end{equation}
The lower bound corresponds to the onset of population inversion near
the Floquet resonance, while the upper bound corresponds to the ideal
limit \(F\to 0\), \CYL{where the Floquet bands take on a band insulator-like filling}. In addition, phonon-mediated relaxation must be below a
critical rate, \(R<R_c\), where
\begin{equation}
\frac{R_c}{M} = \frac{x(3+x^2)}{(1-x)^2},
\quad
x = \sqrt{2\Gamma/M - 1}.
\end{equation}
In the regime \(1/2<\Gamma/M\leq 1\) and \(R<R_c\), the system supports
a finite cavity field, which generates a Floquet gap \(\Delta_F\). The
size of this gap can be estimated by solving Eq.~(\ref{eq:ssfi})
numerically for \(\Delta_F\), and the resulting value is plotted in
Fig.~\ref{fig3}(c).

A key result of the phenomenological model is the \CYNN{tendency} towards an
almost fully occupied lower Floquet band, i.e., \(F_i,F_o\to 0\), as the
lead tunneling rate approaches the ideal limit \(\Gamma/M\to 1\) \CYL{(and in the limit of a lossless cavity)} [see
Fig.~\ref{fig3}(d)]. This regime is often called the ideal Floquet
topological insulator (FTI), because it exhibits a quantized Hall
response of the Floquet bands~\cite{transport}. Since \(\Gamma\) is
controlled by the applied voltage bias \(V_z\), increasing \(V_z\) can
drive the system toward this idealized regime, which explains the
increase in \(\Delta\sigma_{xy}\) with \(V_z\) first presented in
Fig.~\ref{fig1}(c). 
By contrast, close to the threshold
\(\Gamma=M/2\), corresponding to \(|x|\ll 1\), the steady state is
sensitive to phonon-mediated relaxation processes, which strongly
constrain the allowed parameter window to \(R<R_c\).

\CYN{It is worth emphasizing that the result in the limit $\Gamma/M\to 1$ \CYNN{links} gain saturation \CYNN{to the emergence of finite anomalous Hall conductivity}. At small cavity field, the voltage bias creates population inversion outside the resonance ring, populating $S_o$ so that photon-emitting Floquet--Umklapp processes [blue arrow, Figs.~\ref{fig2}(a) and \ref{fig3}(a)] dominate. As the cavity field grows, the Floquet gap increases and this same photon-assisted relaxation processes that produces gain also depletes the population in $S_o$. Saturation therefore occurs as $S_o$ is depleted. Remarkably, this is the same mechanism that gives rise to the ideal FTI state. In particular, in the folded Floquet-band picture, removing the population in $S_o$ is precisely the process that drives the distribution toward a filled lower Floquet band and an empty upper Floquet band. Thus the laser saturation mechanism itself drives the approach to the ideal FTI distribution.}\footnote{\CYNN{We note that the limit \(\Gamma/M\to1\) is singular \CYL{because} \(\Delta_F/\hbar\omega_c\to\infty \) within the perfect-cavity assumption of the phenomenological model, lying outside the perturbative regime in which
the photon-assisted rates scale as \(R(\Delta_F/\hbar\omega_c)^2\). We use this limit only to expose the tendency of the saturation mechanism to drive towards the ideal FTI distribution. When considering finite cavity loss $\alpha \neq 0$, the steady state field amplitude is instead fixed by
\(G(n)=\alpha n\), which cuts off this divergence and yields finite gaps within the controlled regime \(\Delta_F/\hbar\omega_c\ll1\). Our analysis of the microscopic model in Sec.~\ref{sec:micmodel} considers these effects of cavity loss.}}

Despite its simplicity, this phenomenological model captures the
essential structure of the self-organized Floquet steady state. In the
following sections, we show how these features emerge quantitatively
from the microscopic Floquet--Boltzmann description and how they give
rise to measurable transport and geometric signatures.

\begin{figure}[t!]
\includegraphics[width=1\columnwidth]{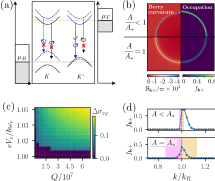}
\caption[]{\textbf{Microscopic model and Hall response.}
(a) Band structure of a magnetically doped TMD with spin-orbit coupling. The top and bottom filtered leads, with chemical potentials $\mu_T$ and $\mu_B$, create a nonequilibrium population. Due to spin-valley splitting, optical selection rules, Pauli blocking, and energy conservation, only one resonant spin-valley transition contributes to gain, selecting a single circularly polarized cavity mode.
(b) Berry curvature $\mathcal{B}_{\mathbf{k}+}$ and occupation $f_{\mathbf{k}+}$ of the upper Floquet band for a field amplitude below the self-consistent steady-state value, $A<A_*$, and at the steady state, $A=A_*$, where $A_*$ denotes the steady-state cavity-field amplitude.
(c) Photo-induced Hall response $\Delta\sigma_{xy}$ as a function of cavity quality factor $Q$ and bias voltage $V_z$.
(d) Momentum cross section of the upper-band occupation near the resonance ring. The shaded regions indicate the momentum windows corresponding to the patches $S_i$ and $S_o$ used in the phenomenological model. At the steady state, the occupation near the resonance is suppressed, consistent with the phenomenological picture.}
\label{fig4}
\end{figure}

\section{Microscopic model for the self-consistent steady state} \label{sec:micmodel}

In this section, we introduce a microscopic model for the coupled electron-photon system and formulate a kinetic description that allows us to determine the nonequilibrium steady state self-consistently. Our goal is to numerically compute the steady-state electronic distribution in the Floquet bands, the resulting photon emission and absorption rates, and the cavity field amplitude selected by the coupled dynamics.

\subsection{Electronic structure and valley selectivity}
We consider a TMD with strong spin orbit coupling and magnetic impurities 
\cite{PhysRevLett.108.196802,PhysRevB.89.155429,PhysRevMaterials.4.104005} that breaks time-reversal symmetry and 
produces a valley and spin-dependent energy shift, such that only {transitions within} one valley and electronic spin 
 {can be} resonant with the cavity frequency, while the {gaps for the opposite spin and/or} valley {are larger and thus detuned from the cavity} [see Fig.~\ref{fig4}(a) for illustration].
As a result, the low-energy dynamics can be restricted to a single valley that is resonantly coupled to the cavity mode. 

To show how weak time-reversal-symmetry breaking selects the active valley, we model the semiconductor using a minimal massive Dirac Hamiltonian
\begin{equation} \label{eq:eqhamil}
\begin{split}
    {H}_{0,\text{el}}(\bk) = \ &\hbar v (k_x \sigma_x +\tau_z k_y\sigma_y)+ \\
    &+ \frac{1}{2} ( \Delta -\frac{\kappa}{2} \tau_z s_z)  \sigma_z + (\Delta_Z + \frac{\kappa}{2} \tau_z) s_z .
\end{split}
\end{equation}
This Hamiltonian describes the conduction and valence bands while capturing the essential Berry-curvature structure of the TMD valleys \cite{PhysRevB.89.155429}. Here, $\mathbf{k}=(k_x,k_y)$ denotes the in-plane crystal momentum, $\tau_z=\pm 1$ labels the $K$ and $K'$ valleys, $s_z=\pm 1$ labels the electronic spin, and $\boldsymbol{\sigma}=(\sigma_x,\sigma_y,\sigma_z)$ is the Pauli vector acting in the pseudospin subspace. The parameter $\kappa$ denotes the spin-orbit-induced energy splitting, while $\Delta_Z$ represents the {effective} \CY{exchange splitting}
due to magnetic doping and $\Delta$ is the direct band gap in the absence of spin-orbit coupling. 
Together, these terms lift the spin and valley degeneracies. The resulting band structure and optical selection rules are illustrated in Fig.~\ref{fig4}(a). \CY{In what follows, we use $\hat c^{\dagger}_{\mathbf{k}\sigma}$ ($\hat c_{\mathbf{k}\sigma}$) to denote the creation (annihilation) operators of the electronic states $|\mathbf{k}\sigma\rangle$ of $H_{0,\mathrm{el}}(\mathbf{k})$ with \CY{2D} momentum $\mathbf{k}$ and band index $\sigma$.} 

{Importantly, in the regime} $\Delta-\kappa/2<\hbar\omega_c <\Delta +\kappa/2$, only a single spin species in each valley is resonant {with the cavity}.
{Moreover}, 
the resonant optical transition within the $K$ valley is entirely Pauli-blocked when $\hbar\omega_c<eV_z< \hbar\omega_c + 2\Delta_Z$, leaving only a single optical transition in the valley $K'$. \CY{Here, $eV_z = \mu_T - \mu_B$ denotes the difference in the chemical potentials of the top and bottom filtered leads, see Sec.~\ref{sec:elpopdyn} for further discussion, which are chosen so that the optically-active spin species in the $K'$ valley is charge neutral.}
Suitable TMD materials include $\mathrm{MoS_2}$ and $\mathrm{WSe_2}$, which exhibit strong spin-orbit coupling reaching $\kappa \sim 0.1-0.4 \ \mathrm{eV}$ and large \CY{exchange} splitting $\Delta_Z \sim 0.1 \ \mathrm{eV}$ under magnetic doping \cite{PhysRevB.89.155429}.

{Focusing on the single resonant mode {and electronic states near the band edges}, the chiral}-cavity field couples to interband transitions via minimal coupling [see Eq.~\eqref{eq:effective_floquet_ham}] {described by}: 
\begin{equation}
\hat{O}_{e-p} =   \int d^2\boldsymbol{k}/(2\pi)^2 \hat{\boldsymbol{\psi}}_{\boldsymbol{k}}^\dagger {O}_{e-p} \hat{\boldsymbol{\psi}}_{\boldsymbol{k}},\quad {O}_{e-p}=\lambda\,\sigma_+,
\end{equation} 
where $\sigma_\pm=(\sigma_x\pm i\sigma_y)/2$ denote circularly polarized interband operators and $\lambda = ev \mathcal{E}_0/\omega_c$ denotes the light--matter coupling strength {[see Eq.~(\ref{eq:field_strength}) for definition of the characteristic field amplitude of the cavity, $\mathcal{E}_0$]}. This coupling preserves crystal momentum and selectively addresses a single valley, consistent with circular dichroism in TMD monolayers.
The resulting Floquet Hamiltonian hybridizes the valence band with the one-photon replica of the conduction band near the resonance condition, opening a Floquet gap
\begin{equation} \label{eq:deltaf}
    \Delta_F = 2 \lambda |A| \cos^2(\theta/2) ,
\end{equation}
where $\tan(\theta) =\sqrt{(\hbar\omega_c)^2 - (\Delta-\kappa/2)^2}/(\Delta-\kappa/2)$, along a closed resonance ring in momentum space, see Figs.~\ref{fig2}(a) and (b).

\CYN{The single-helicity approximation is accurate when choosing the cavity
frequency $\hbar\omega_c$ close to the active optical gap $\tilde \Delta \equiv \Delta - \kappa /2$, corresponding to small $\theta$. For the selected circular
polarization, the resonant matrix element opens the Floquet gap, given by Eq.~(\ref{eq:deltaf}), whereas the opposite circular polarization would open the smaller gap scale $\Delta_F^{\rm opp} = 2\lambda |A|\sin^2(\theta/2)$, since $\theta$ is small. Because the stimulated-emission rate is proportional to the square of the Floquet-Umklapp matrix element, the small-signal gain of the opposite helicity is suppressed by $\sim \tan^4(\theta/2)$. Taking
$\hbar\omega_c-\tilde \Delta\ll \tilde \Delta$, one can approximate $\tan^4(\theta/2) \approx \frac{1}{4} ( {\hbar\omega_c}/{\tilde \Delta}-1)^2$.  Thus, when the cavity frequency is chosen only slightly above the active band edge, one helicity mode survives, while the opposite helicity mode remains parametrically suppressed.}

\subsection{Coupling to leads}

The leads above and below the TMD are treated as macroscopic reservoirs characterized by chemical potentials $\mu_T$ and $\mu_B$, respectively, with each held at thermal equilibrium at zero temperature \CY{and filtered so that the contacts have density of states in non-overlapping energy windows, see Sec.~\ref{sec:elpopdyn} for more details.} 
The coupling between the electronic system and the leads is described by a tunneling Hamiltonian,
\begin{equation}
\hat H_{\mathrm{leads}} = \sum_{\ell=T,B} \sum_{\mathbf{p}} \varepsilon_{\ell \mathbf{p}}\,
\hat d^{\dagger}_{\ell \mathbf{p}} \hat d_{\ell \mathbf{p}},
\end{equation}
\begin{equation}
\hat H_{\mathrm{tun}} =
\sum_{\ell=T,B} \sum_{\mathbf{p},\mathbf{k},\sigma}
\left(
t_{\ell \mathbf{p} \mathbf{k} \sigma}\,
\hat d^{\dagger}_{\ell \mathbf{p}} \hat c_{\mathbf{k}\sigma}
+ \mathrm{h.c.}
\right),
\label{eq:tun}
\end{equation}
where $\hat d^{\dagger}_{\ell\mathbf{ p}}$ ($\hat d_{\ell \mathbf{p}}$) creates (annihilates) an electron with momentum $\mathbf{p}$ in \CY{the three-dimensional} lead $\ell$ \cite{transport}. The tunneling amplitudes $t_{\ell \mathbf{p} \mathbf{k} \sigma}$ determine the strength of the coupling to the leads, \CY{and we use $eV_z = \mu_T - \mu_B$ to denote the voltage bias.}

\subsection{Electron--phonon coupling}

The phonon environment is modeled as a bath of acoustic phonons with linear dispersion {and optical phonons with zero (flat) dispersion},
\begin{equation}
\hat H_{\mathrm{ph}} =
\sum_{\mathbf{q},\zeta=a,o} \hbar \omega_{\mathbf{q}}^\zeta \, \hat b^\dagger_{\mathbf{q}} \hat b_{\mathbf{q}},
\qquad
\omega_{\mathbf{q}}^a = v_s |\mathbf{q}|, \quad \omega_{\mathbf{q}}^o = \omega_o,
\label{eq:PhononDispersion}
\end{equation}
where $\hat b^{\zeta \dagger}_{\mathbf{q}}$ ($\hat b_{\mathbf{q}}^\zeta$) creates (annihilates) an acoustic ($\zeta = a$) or optical ($\zeta = o$) phonon with {in-plane} momentum $\mathbf{q}$, $v_s$ denotes the speed of sound of the acoustic phonon branch, and $\omega_o$ models the approximately-dispersionless optical phonon mode. The phonon bath is assumed to remain in thermal equilibrium at zero temperature.

The electron--phonon interaction is described by the Hamiltonian
$\hat H_{\mathrm{el\text{-}ph}} = \sum_{\zeta} \hat H_{\mathrm{el\text{-}ph}}^{\zeta}$, where
\begin{equation}
\hat H_{\mathrm{el\text{-}ph}}^\zeta  =
\sum_{\mathbf{k},\mathbf{q},\sigma}
g_{\mathbf{q}}^\zeta \,\hat c^\dagger_{\mathbf{k}+\mathbf{q},\sigma}
\, \hat c_{\mathbf{k},\sigma}
\left(
\hat b_{\mathbf{q}}^\zeta + \hat b^{\zeta \dagger}_{-\mathbf{q}}
\right).
\end{equation}
where the coupling matrix element for the phonon mode indexed by $\zeta$ is denoted $g_{\mathbf{q}}^{\zeta}$. \CY{In our simulations, we include an effective coupling to the 2D longitudinal acoustic and optical phonon modes of the TMD, as described in detail in the Supplemental Information (SI).}

The electronic steady state is determined by solving a set of kinetic equations for the occupations of the Floquet quasistates, which account for particle injection and extraction by the leads, energy relaxation via phonon emission, and photon-assisted transitions mediated by the cavity field. These equations are supplemented by a rate equation for the cavity photon number [Eq.~\eqref{eq:rate_eq}], which captures the balance between photon emission and cavity losses. 

\subsection{Rate equations for electronic occupations and cavity-photon gain}

The combined coupling to the leads and to the phonon bath drives the electronic system into a nonequilibrium steady state that is most naturally described at the level of the occupations of Floquet quasistates. 
The corresponding Floquet occupation functions are
\begin{equation}
f_{\mathbf{k}\nu}(t)
=
\left\langle
\hat \phi^\dagger_{\mathbf{k}\nu}(t)\,
\hat \phi_{\mathbf{k}\nu}(t)
\right\rangle ,
\label{eq:floquet_occupation}
\end{equation}
where we defined fermionic operators $\hat \phi_{\mathbf{k}\nu}(t)$ that annihilate an electron in the Floquet state $|\phi_{\mathbf{k}\nu}(t)\rangle$.

In the weak-coupling limit to the leads and phonon bath, the time evolution of these occupations is governed by a kinetic (Floquet--Boltzmann) equation \cite{PhysRevA.92.062108,PhysRevA.91.033601,fbe_orig,fbe_adv,Parmee2020,PhysRevResearch.3.L012016},
\begin{equation}
\dot f_{\mathbf{k}\nu}
=
\sum_{\mathbf{k}'\nu'}
\Big(
W_{\mathbf{k}'\nu'\rightarrow\mathbf{k}\nu}\,
-
W_{\mathbf{k}\nu\rightarrow\mathbf{k}'\nu'}
\Big)
+
\Gamma^{\mathrm{lead}}_{\mathbf{k}\nu},
\label{eq:floquet_kinetic}
\end{equation}
where
\begin{equation}
W_{\mathbf{k}'\nu'\rightarrow\mathbf{k}\nu}
= \sum_s
R^s_{\mathbf{k}\nu,\mathbf{k}'\nu'}\,
f_{\mathbf{k}'\nu'} (1-f_{\mathbf{k}\nu})
\end{equation}
and $R^s_{\mathbf{k}\nu,\mathbf{k}'\nu'} = R^{s,\mathrm{ph}}_{\mathbf{k}\nu,\mathbf{k}'\nu'} + R^{s,\mathrm{rec}}_{\mathbf{k}\nu,\mathbf{k}\nu'} \delta_{\mathbf{k}\mathbf{k}'}$ denotes the intrinsic transition rate from Floquet state $(\mathbf{k}',\nu')$ to $(\mathbf{k},\nu)$, split into transition rates induced by electron-phonon collisions $R^{s,\mathrm{ph}}_{\mathbf{k}\nu,\mathbf{k}'\nu'}$, and by vertical radiative recombination processes $R^{s,\mathrm{rec}}_{\mathbf{k}\nu,\mathbf{k}\nu'}$. 

The contribution to the electronic kinetics arising from tunneling to the leads is described by $\Gamma^{\mathrm{lead}}_{\mathbf{k}\nu}
=\Gamma^{\mathrm{lead}}_{
T\mathbf{k}\nu}+\Gamma^{\mathrm{lead}}_{
B\mathbf{k}\nu}$, where 
\begin{equation}
\Gamma^{\mathrm{lead}}_{
\ell\mathbf{k}\nu}
= \sum_s
\Gamma^s_{\ell,\mathbf{k}\nu}
\left[
f_\ell(\varepsilon_{\mathbf{k}\nu}+s\hbar\omega_c)
-
f_{\mathbf{k}\nu}
\right].
\label{eq:lead_term}
\end{equation}
Here, $\Gamma^s_{\ell,\mathbf{k}\nu}$ denotes the tunneling rate between the Floquet state with quasienergy $\varepsilon_{\mathbf{k}\nu}+s\hbar\omega_c$ and top/bottom lead $\ell=T,B$, where $f_\ell(\varepsilon)$ is the Fermi distribution of lead $\ell$. 
In the steady state, the occupations satisfy $\dot f_{\mathbf{k}\nu}=0$, yielding a generally nonthermal distribution that depends parametrically on the cavity field amplitude. Microscopically, the transition rates can be approximated using Fermi's golden rule modified for Floquet states \cite{fbe_orig}, the details of which we provide in \CY{Supplemental} Sec.~\ref{sec:numsol}. \CY{Assuming filtered leads with a constant density of states, the lead tunneling rate simplifies to $\Gamma^s_{\ell,\mathbf{k}\nu} = \Gamma^{\text{lead}} \langle \phi^s_{\mathbf{k}\nu} | \phi^s_{\mathbf{k}\nu} \rangle$, where $\Gamma^{\text{lead}}$ is a constant effective lead tunneling rate.} 

\IE{In order to determine the gain function \(G(n)\), we compute the rate at which electronic transitions change the photon content of the dressed steady state. Since the electronic states are Floquet dressed by the coherent cavity field, each state \((\mathbf{k},\nu)\) carries an average Floquet sideband index \cite{transport,fbe_adv,fbe_orig,floquethandbook}
\begin{equation}
\bar m_{\mathbf{k}\nu}
=
\sum_m m \,
\langle \phi_{\mathbf{k}\nu}^{m} | \phi_{\mathbf{k}\nu}^{m} \rangle ,
\label{eq:photon_content}
\end{equation}
where
$|\phi_{\mathbf{k}\nu}(t) \rangle
=
\sum_m |\phi_{\mathbf{k}\nu}^{m} \rangle e^{-im\omega_c t}$.
Thus, a transition from \((\mathbf{k}',\nu')\) to \((\mathbf{k},\nu)\) changes the cavity photon balance by the difference between the final and initial sideband contents, together with the sideband shift \(s\). Summing all such transitions gives
\begin{equation}
\begin{split}
G(n)
&=
\sum_{\mathbf{k}\nu,\mathbf{k}'\nu',s}
R_{\mathbf{k}\nu,\mathbf{k}'\nu'}^s
\left(\bar m_{\mathbf{k}\nu}-\bar m_{\mathbf{k}'\nu'} + s\right)
f_{\mathbf{k}'\nu'}\bigl(1-f_{\mathbf{k}\nu}\bigr) \\
&+ \sum_{\mathbf{k}\nu,s,\ell} \Gamma^{s}_{\ell\mathbf{k}\nu}
\left(\bar m_{\mathbf{k}\nu}+s\right) \left[
f_\ell(\varepsilon_{\mathbf{k}\nu}+s\hbar\omega_c)
-
f_{\mathbf{k}\nu}
\right].
\end{split}
\label{eq:photon_rate}
\end{equation}}

\subsection{Microscopic origin of gain saturation}

The dependence of the photon emission rate on the cavity occupation is shown in Fig.~\ref{fig1}(b). The solid curves represent the electronic contribution $G(n)$ computed from the microscopic Floquet--Boltzmann equations for several bias voltages $V_z$ \CY{[which enters via the lead occupation function $f_{\ell} (\varepsilon)$ defined in Eq.~(\ref{eq:lead_term})]} slightly exceeding the photon energy $\hbar\omega_c$. 
For small photon number, emission 
{into the cavity} dominates absorption due to population inversion near the Floquet resonance, leading to positive gain. As $n$ increases, the Floquet gap $\Delta_F$ grows, modifying both the Floquet spectrum and the steady-state occupations. This feedback suppresses further emission and ultimately leads to saturation.

The dashed curve in Fig.~\ref{fig1}(b) shows the cavity loss rate $\alpha n$, which increases linearly with photon number. The steady-state photon occupation is determined by the intersection of gain and loss curves, indicated by the markers. Increasing the bias voltage \CY{$V_z$}
enhances population inversion and shifts the intersection to larger $n$, resulting in a larger self-organized Floquet gap.

\CYN{The steady state discussed above corresponds to a stable \CYNN{photon population} for the cavity mode. This can be seen by linearizing the photon rate equation, Eq.~(\ref{eq:rate_eq}), around the steady state, \(n=n^\ast+\delta n\), which gives
\[
    \delta \dot n =
   \left. \frac{\partial \dot {n}}{\partial n} \right|_{n=n^*}\delta n .
\]
Thus, the finite-field steady state solution is stable when the net photon-production rate \(\dot{n} = G(n)-\alpha n\)
crosses zero from positive to negative as \(n\) is increased. This is
the case for the steady state solutions shown in Fig.~\ref{fig1}(b): the gain exceeds the cavity loss for \(n<n^*\), while it falls below
the loss for \(n>n^*\). Small deviations from \(n^*\) therefore
decay in time, so the self-consistent cavity field corresponds to a
stable fixed point.}

The 
steady state Floquet gap $\Delta_F$ is summarized in Fig.~\ref{fig2}(c) as a function of the cavity quality factor $Q$ and the applied bias $V_z$ across the TMD. A finite gap develops only when $eV_z >\hbar\omega_c$, reflecting the threshold condition for photon emission. Higher $Q$ factors allow larger steady-state fields and hence larger Floquet gaps.

\section{Signatures of the Floquet steady state} \label{sec:signaturesss}

The self-organized Floquet steady state gives rise to a set of distinct and experimentally accessible signatures. These signatures reflect the emergence of a coherent periodic drive, the opening of a Floquet gap, and the associated geometric and transport properties of the dressed electronic bands. Importantly, several of these effects can be probed using dc or steady-state measurements, without the need for ultrafast optical techniques.

\begin{figure}[t]
\includegraphics[width=1\columnwidth]{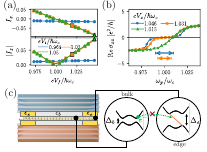}
\caption[]{\textbf{Experimental signatures of the Floquet gap.}
(a) Out-of-plane current response following a bias quench. The cavity is initialized at bias $V_i$ and rapidly quenched to $V_f$, while the field amplitude remains approximately fixed during the measurement. The plotted current is normalized by $e\Gamma^{\mathrm{lead}}$, and the suppression of $|I_z|$ near $eV_f\simeq\hbar\omega_c$ signals the Floquet gap.
(b) Optical conductivity $\sigma_{xx}(\omega_p)$ as a function of probe frequency $\omega_p$ for different biases $V_z$. The colored arrows indicate the frequency window associated with the Floquet gap.
(c) Proposed setup for probing edge transport. Dielectric engineering enhances the cavity field near the edge, producing an edge gap $\Delta_e$ larger than the bulk gap $\Delta_b$. \IE{If bulk to edge scattering is sufficiently slow and suitable edge contacts can be implemented, this geometry may allow future probes of edge state transport.}
}
\label{fig5}
\end{figure}

\subsection{Berry curvature and in-plane anomalous Hall response}

\IE{The material model introduced in Eq.~\eqref{eq:eqhamil} explicitly breaks time reversal symmetry, as appropriate for a TMD with dilute magnetic impurities. As a result, the undriven bands already possess a finite Berry curvature and can support a background anomalous Hall response, depending on the steady state occupation. Thus, in our setup the role of the cavity field is to introduce an additional Floquet contribution by resonantly hybridizing conduction and valence band replicas.}

{In Fig.~\ref{fig4}(b) we show} the \CY{time-averaged} and momentum-resolved 
Berry curvature of the upper Floquet band together with the corresponding steady-state occupation. The Berry curvature is sharply peaked along the resonance ring, reflecting the avoided crossing between Floquet replicas of the conduction and valence bands. At the same time, dc pumping and phonon relaxation lead to a pronounced nonequilibrium occupation imbalance across the gap.

\IE{To isolate the contribution arising from cavity-induced Floquet dressing, we focus on the change in Hall conductivity, \(\Delta\sigma_{xy}\), relative to the undriven value, as shown in Fig.~\ref{fig4}(c). } Consistent with the phenomenological model [see Fig.~\ref{fig3}(b)], the enhancement of \(\Delta\sigma_{xy}\) is controlled by the steady state occupation \(f_{\boldsymbol{k}+}\) of the upper Floquet band. As the cavity field amplitude grows, this occupation is depleted near the resonance ring, Fig.~\ref{fig4}(d), leading to an enhanced Hall response and a clear transport signature of the self organized Floquet steady state.

\subsection{Out-of-plane current response to a bias quench}

Beyond in-plane responses, the opening of a Floquet gap also leads to sharp signatures in the out-of-plane transport {through the contacts used to drive the system}.
In particular, the gap strongly affects charge transport through the out-of-plane contacts that inject and extract carriers from the active material. One way to probe this effect is through a rapid quench of the bias voltage from $V_i$ to $V_f$, such that the initial value,  $V_i$, is above the lasing threshold. 
If \(V_f\) is below the lasing threshold, the gain no longer compensates cavity loss and the coherent field decays with the cavity photon lifetime, $\tau_{\rm cav} \approx Q/\omega_c$.
For a high-quality cavity with \(Q\sim 10^6\)  
and near infrared frequency, \CY{this lifetime can reach hundreds of picoseconds to nanoseconds, corresponding to a decay rate in the GHz range.  Thus, the relevant experimental requirement is electrical modulation on the GHz scale, which is much slower than the electronic response time of the leads.}

The out-of-plane current can be expressed directly in terms of the tunneling rates entering the Floquet--Boltzmann equation. The steady-state current is given by
\begin{equation}
I_z
=
e
\sum_{\mathbf{k}\nu}
\left(
\Gamma^{\mathrm{lead}}_{T,\mathbf{k}\nu}
-
\Gamma^{\mathrm{lead}}_{B,\mathbf{k}\nu}
\right),
\end{equation}
where $\Gamma^{\mathrm{lead}}_{\ell,\mathbf{k}\nu}$ denotes the net tunneling rate between Floquet state $(\mathbf{k},\nu)$ and lead $\ell$, as defined in Eq.~\eqref{eq:lead_term}. This expression automatically incorporates Pauli blocking and includes both elastic and Floquet-assisted tunneling processes. The current as a function of the quench bias voltage $V_f$ is shown in Fig.~\ref{fig5}(a). 
The current is suppressed when the quenched bias $V_f$ is tuned so that the chemical potentials of the leads lie within the Floquet gap. This suppression occurs because the density of available Floquet states at the relevant quasienergies is strongly reduced, making inelastic processes necessary to bridge the gap. Consequently, the out-of-plane current becomes highly sensitive to bias near the Floquet gap. Moreover, the current changes sign as $V_f$ is tuned across the gap. 
For $eV_f<\hbar\omega_c$, photo-induced population excitation drives a reverse current through the leads.

\subsection{Optical signatures of the Floquet gap}

The Floquet gap size can also be measured through the optical response of the TMD. A probe beam with frequency $\omega_p$ can penetrate the contacts by propagating at a slightly non-orthogonal angle relative to the Bragg reflectors. As a function of $\omega_p$, the optical conductivity 
\begin{equation}
    \sigma_{xx}(\omega_p) = \frac{i}{\omega_p}  \frac{e^2}{4\pi^2} \sum_{\mathbf{k}} \sum_{\alpha,\gamma,m} \frac{(f_{\mathbf{k}\alpha} - f_{\mathbf{k}\gamma})
  |V_{\alpha\gamma,m}(\mathbf{k})|^2}{\hbar\omega_p + \varepsilon_{\mathbf{k}\alpha} - \varepsilon_{\mathbf{k}\gamma} - m\hbar\omega_c + i\eta}
\end{equation}
\CY{is calculated using the steady state occupations $f_{\mathbf{k}\alpha}$ derived from the microscopic TMD model}. Here,
\begin{equation}
    V_{\alpha\gamma,m}(\mathbf{k}) = \sum_l \langle \phi_{\mathbf{k}\alpha}^{l}| v \sigma_x| \phi_{\mathbf{k}\gamma}^{l+m} \rangle,
\end{equation}
and $\eta = 0^+$. The real part of the conductivity, shown in Fig.~\ref{fig5}(b),
develops a plateau at $\mathrm{Re} \ \sigma_{xx}=0$ whenever the probe frequency lies within the Floquet gap. This zero-conductivity plateau separates a regime of positive optical conductivity for $\omega_p>\omega_c$, where the probe induces absorption between filled valence-band states and empty conduction-band states, from a regime of negative optical conductivity for $\omega_p<\omega_c$, where the probe is resonant with the population-inverted states. For reference, the gap size is marked by arrows, which matches closely with the frequency window for the plateau. In this frequency window, the TMD becomes transparent because there are no available electronic states for optical absorption. This optical feature therefore provides a direct experimental signature of the Floquet gap and its magnitude.

\subsection{Signatures of the edge states}
Another experimentally accessible signature can arise from edge-current measurements. In the proposed setup, shown in Fig.~\ref{fig5}(c), conducting materials with dielectric constants $\epsilon_b$ and $\epsilon_e$ surround the bulk and near the edge of the TMD, respectively. When $\epsilon_e < \epsilon_b$, the cavity field is enhanced near the edge, resulting in a larger Floquet gap $\Delta_e$ in this region (see inset). If the electron scattering time from the bulk to the edge, $\tau_{\text{eb}}$, is sufficiently long to satisfy $\hbar/\Delta_e \ll \tau_{\text{eb}}$, \CY{and the inelastic Floquet-Umklapp scattering time within the edges, $\tau_{\mathrm{edge}}^{\mathrm{FU}}$ is much longer than the transit time of a single electron around the edges of the sample, $\sim L/v_e$,}
then the edge states can be biased independently through narrow-band, filtered edge contacts and can support quantized conductivity \cite{transport,PhysRevB.93.045121,PhysRevB.110.075428,PhysRevLett.115.106403}. \CY{Here, $L$ denotes the edge length of the sample, and $v_e \sim \Delta_F/(\hbar k_R)$ denotes the velocity of the edge state, where $k_R$ is the magnitude of the electronic momentum along the single-photon resonance.}
For infrared frequencies, the Floquet gap [see Fig.~\ref{fig2}(c)] can exceed $O(10 \ \mathrm{meV})$, corresponding to $\hbar/\Delta_e \sim O(10 \ \mathrm{fs})$. This timescale is much shorter than the phonon-mediated bulk-to-edge scattering time, which typically occurs on timescales of $\tau_{\text{eb}} \sim 1$--$10 \ \mathrm{ps}$. \CY{Additionally, for sample sizes on the order $L \sim 1 \ \mathrm{\mu m}$ and typical edge state velocities $v_e \sim 10^5 \ \mathrm{m/s}$ (where we have assumed $k_R \sim 10^8 \ \mathrm{m}^{-1}$), the transit time of the electrons around the sample edge, $L/v_e \sim 10 \ \mathrm{ps}$, can be much smaller than the timescale for phonon-mediated Floquet-Umklapp processes, $\tau_{\mathrm{edge}}^{\mathrm{FU}} \sim (1 \ \mathrm{ps}) (\hbar\Omega/\Delta_e)^2 \sim 10 \ \mathrm{ns}$.}

\section{Classical characteristics and power balance}

\CYN{We now address the classical operating constraints of the proposed device. The central questions are: how much electrical power is required to sustain the self-organized Floquet field, what fraction of that power is converted into coherent cavity photons, and how much heat is generated by the active material and contacts. These considerations expose the advantage of the cavity scheme: not only that it generates a large photon field, but that it does so efficiently through a gain-loss balance that utilizes less power than injecting the full optical field intensity from outside the device.}

\begin{figure}[t]
\includegraphics[width=1\columnwidth]{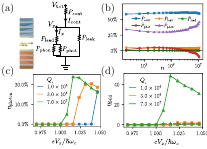}
\caption[]{\textbf{Power budget and energy efficiency.}
(a) Effective circuit model for the electrically driven cavity material device. The applied voltage $V_{\mathrm{tot}}$ drives a current $I_{\mathrm{cont}}$ supplied through the contacts, which splits into the vertical current through the TMD, $I_z$, and a leakage current bypassing the active region.
(b) Power budget as a function of the steady-state photon occupation $n$. The total power supplied to the device, $P_{\mathrm{tot}}=I_{\mathrm{cont}}V_{\mathrm{tot}}$, is distributed among photon generation, phonon heating, contact dissipation, leakage dissipation, and lead dissipation.
(c) Photon-generation efficiency $\eta_{\mathrm{phot}}=P_{\mathrm{phot}}/P_{\mathrm{tot}}$ as a function of bias $eV_z$ for several intrinsic cavity quality factors $Q$.
(d) Field conversion ratio $\eta_{\mathrm{field}}=\mathcal{I}_{\mathrm{field}}\mathcal{A}/P_{\mathrm{tot}}$, where $\mathcal{I}_{\mathrm{field}}=\frac{1}{2}c\epsilon_0\mathcal{E}_*^2$ is the steady-state field intensity, $\mathcal{E}_*$ is the steady-state electric field amplitude, and $\mathcal{A}$ is the TMD area.}
\label{fig6}
\end{figure}

\subsection{Power balance in the steady state}

In steady-state operation, the device is powered by a dc bias $V_{\mathrm{tot}}$ applied across the electronic leads, resulting in a total electrical power 
\begin{equation} 
P_{\mathrm{tot}} = I_{\text{cont}} V_{\mathrm{tot}} ,
\end{equation}
where $I_{\mathrm{cont}}$ is the total current flowing between the contacts [see Fig.~\ref{fig6}(a)]. This injected power is redistributed among several distinct channels associated with transport, dissipation, and cavity dynamics.

In particular, the total electrical power $P_{\mathrm{tot}}$ can be decomposed into four contributions,
\begin{equation}
P_{\mathrm{tot}}
=
P_{\mathrm{sys}}
+
P_{\mathrm{lead}}
+
P_{\mathrm{leak}}
+
P_{\mathrm{cont}} 
.
\end{equation}

The first contribution,
\begin{equation}
    P_{\mathrm{sys}} = P_{\mathrm{phon}} + P_{\mathrm{phot}}
\end{equation}
is the power dissipated by the active TMD material. It includes the power $P_{\mathrm{phon}}$ dissipated into lattice vibrations via electron-phonon scattering within the active material, and the power 
\begin{equation}
P_{\mathrm{phot}}
=
\hbar\omega_c\,G(n) ,
\end{equation}
required to sustain the steady-state photon population in the cavity. The latter compensates photon losses and represents the fraction of electrical power converted into coherent electromagnetic radiation. In the steady state, $P_{\mathrm{phot}}$ balances the intrinsic cavity loss rate. The total power $P_{\mathrm{sys}}$ is defined microscopically from the tunneling processes entering the Floquet--Boltzmann equation as \begin{equation}
P_{\mathrm{sys}}
=
\sum_{\mathbf{k}\nu,\ell,s}
(\varepsilon_{\mathbf{k}\nu}+s\hbar\omega_c)\,
\Gamma^s_{\ell,\mathbf{k}\nu}
\left[
f_\ell(\varepsilon_{\mathbf{k}\nu}+s\hbar\omega_c)
-
f_{\mathbf{k}\nu}
\right].
\end{equation}

The second contribution $P_{\mathrm{lead}}$
corresponds to the power dissipated by the leads connecting the doped distributed Bragg reflectors to the TMD. This term is defined as 
\begin{equation}
    P_{\mathrm{lead}} = I_z V_z - P_{\mathrm{sys}}.
\end{equation}

The third contribution,
\begin{equation}
    P_{\mathrm{leak}} = (I_{\mathrm{cont}} -I_z) V_z
\end{equation}
is the power that is dissipated by leakage current that corresponds to a direct tunneling between the leads, therefore bypassing the TMD.

Finally, the remaining power $P_{\mathrm{cont}}$ is dissipated at the contacts. Physically, this contribution accounts for Joule heating associated with charge relaxation in the leads and contact regions. 

Fig.~\ref{fig6}(b) 
illustrates this steady-state power balance by showing the injected power, the power transferred into the cavity mode, and the power dissipated into phonons as a function of the cavity photon occupation. The decomposition highlights that a finite fraction of the electrical power is coherently redirected into the cavity field, while the remaining dissipation is shared between phonon heating and contact losses.

\subsection{Efficiency of the steady-state operation}
 In Fig.~\ref{fig6}(c), we demonstrate the photon-conversion efficiency, defined as
\begin{equation}
\eta_{\mathrm{phot}} = \frac{P_{\mathrm{phot}}}{P_{\mathrm{tot}}}.
\end{equation}
A sharp increase appears once $V_z$ is tuned above $\hbar\omega_c$, corresponding to the regime in which the lead bias produces a substantial rate of stimulated emission. At higher \CY{bias $V_z$}, however, this efficiency is suppressed because the larger current flowing through the system causes dissipation in the contact resistance to dominate. 

Lastly, to compare the cavity-field intensity with the total input power, we also compute \CYNN{the ratio}
\begin{equation}
\eta_{\mathrm{field}} = \frac{\mathcal{I}_{\mathrm{field}} \mathcal{A}}{P_{\mathrm{tot}}},
\end{equation}
where
\begin{equation}
\mathcal{I}_{\mathrm{field}} = \frac{1}{2} c \epsilon_0 \mathcal{E}_*^2
\end{equation}
is the field intensity, $\mathcal{A}$ is the total surface area of the TMD, $\mathcal{E}_*$ is the steady-state electric-field amplitude, $c$ is the speed of light. Fig.~\ref{fig6}(d) demonstrates that the system can be tuned into a regime where $\eta_{\mathrm{field}}\gg 1$ in an efficient cavity. For comparison, an equivalent field intensity generated by direct laser irradiation at the same contact bias, leads to a \CYNN{ratio} of
\begin{equation}
\tilde \eta_{\mathrm{field}} = \frac{\mathcal{I}_{\mathrm{field}}\mathcal{A} }{\mathcal{I}_{\mathrm{field}} \mathcal{A} + P_{\mathrm{tot}}} < 1
\end{equation}
\CYNN{due to the additional power $\mathcal{I}_{\mathrm{field}}\mathcal{A}$ needed to generate the field intensity through an external laser source.}

\subsection{Current--voltage characteristics}

At low bias, below the threshold for cavity field buildup $e V_z< \hbar\omega_c$, transport is governed by the equilibrium band structure and conventional inelastic scattering. In this regime, the current increases smoothly with voltage, leading to an approximately linear $I_z$--$V_z$ characteristic, see Fig.~\ref{fig7}(a). 

\CY{Once the bias exceeds the resonance condition $eV_z \simeq \hbar\omega_c$, the cavity develops a finite steady-state field whose amplitude increases with $V_z$. This onset is accompanied by a sharp rise in the current, reflecting a substantial increase in the electrical power supplied by the leads, see Fig.~\ref{fig7}(a). Microscopically, this input power compensates two field-dependent loss channels. The first is photon leakage from the cavity, which scales as $\alpha n_*$, where $n_*$ is the steady-state photon occupation. The second is electronic heating induced by the Floquet drive, which grows as $\sim \Delta_F^2$, as predicted by the phenomenological model. Together, these loss mechanisms require an increase in the power input from the leads that grows sharply with the field amplitude, producing the observed sharp increase in current.}

\subsection{Effective cavity quality factor}

The steady-state photon occupation is determined by the balance between electronic gain and cavity loss, which can be expressed in terms of an effective quality factor $Q_{\mathrm{eff}}$. This quantity captures the combined effect of intrinsic cavity losses and electronic backaction.

\begin{figure}[t]
\includegraphics[width=1\columnwidth]{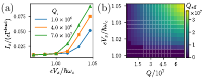}
\caption[]{\textbf{Electrical current and effective cavity quality factor.}
(a) Vertical current through the TMD per unit cell, as a function of bias $eV_z$ for several intrinsic cavity quality factors $Q$. Above the threshold $eV_z>\hbar\omega_c$, the electrically driven population inversion produces stimulated emission into the cavity mode.
(b) Effective cavity quality factor $Q_{\mathrm{eff}}$ as a function of bias $eV_z$ and intrinsic quality factor $Q$. The reduction of $Q_{\mathrm{eff}}$ reflects additional photoabsorption by the TMD.}
\label{fig7}
\end{figure}

Fig.~\ref{fig7}(b) shows the effective cavity quality factor as a function of bias voltage and bare cavity $Q$. Below threshold, $Q_{\mathrm{eff}}$ remains small and the cavity field does not build up. Above threshold, electronic gain compensates cavity losses, leading to a rapid increase in $Q_{\mathrm{eff}}$ and stabilizing a finite photon population. The resulting Floquet gap and associated transport signatures therefore emerge only within a well-defined region of parameter space controlled by voltage and cavity losses.

\subsection{Heating and cooling considerations}

In steady-state operation, electrical power injected into the device ultimately appears as heat that must be removed by the surrounding environment. In the present setup, heating arises from two distinct contributions: power dissipated internally into lattice vibrations, $P_{\mathrm{phon}}$, and power dissipated at the contacts, $P_{\mathrm{cont}}$.

The phonon contribution $P_{\mathrm{phon}}$ represents energy transferred from the driven electronic system to the lattice within the active material via electron--phonon scattering. This channel determines the local heating of the driven region and is therefore the most relevant constraint for maintaining low electronic temperatures in the presence of Floquet band dressing.

In addition, a finite fraction of the total electrical power is dissipated at the contacts,
where charge relaxation occurs in the leads and contact regions rather than in the active material itself. While this contribution does not directly heat the Floquet-engineered region, it nevertheless contributes to the overall thermal load of the device and must be accounted for in practical implementations.

To maintain the device at low temperature, the cooling power supplied by the substrate or cryogenic environment must satisfy
\begin{equation}
P_{\mathrm{cool}} \gtrsim P_{\mathrm{phon}} + P_{\mathrm{cont}} + P_{\mathrm{leak}} + P_{\mathrm{lead}}.
\end{equation}
This requirement is consistent with standard cryogenic cooling capabilities for mesoscopic and nanoscale devices, particularly because strong light--matter coupling allows the Floquet steady state to be established at relatively low injected power. The minimum cooling power, $P_{\mathrm{cool}}^* = P_{\mathrm{phon}} + P_{\mathrm{cont}} + P_{\mathrm{leak}} + P_{\mathrm{lead}}$, may be estimated from 
\begin{equation}
    P_{\mathrm{cool}}^* = \frac{1-\eta_{\mathrm{phot}}}{\eta_{\mathrm{phot}}} P_{\mathrm{phot}},
\end{equation}
with $P_{\mathrm{phot}} = \hbar \omega_c^2 n_*/  Q$, where $n_*$ denotes the steady-state photon occupation. Using representative parameters $\mathcal{V} \sim 10~\mathrm{\mu m}^3$, $Q \sim 10^8$, $n_* \sim 10^6$ [see Fig.~\ref{fig1}(b)], $\eta_{\mathrm{phot}} \sim 0.1$, and $\hbar\omega_c \sim 1~\mathrm{eV}$ in the near-infrared, we estimate 
\CY{$P_{\mathrm{cool}}^* \sim 20 \ \mu\mathrm{W}$}. 

A key advantage of the cavity-mediated Floquet scheme is that increasing the cavity quality factor lowers the threshold for photon buildup and redirects a larger fraction of the injected electrical power into the cavity field. This reduces the power dissipated as heat, both internally and at the contacts, and enables operation in a regime where coherent Floquet band engineering coexists with manageable classical dissipation.

\section{Discussion and outlook}

We have presented a framework for realizing Floquet-\CY{engineering at strong self-generated Floquet fields} in a dc-driven quantum material without external laser illumination. Instead, the periodic drive emerges self-consistently from the coupled dynamics of an electrically pumped electronic system and a cavity mode. The resulting nonequilibrium steady state is characterized by a stable limit cycle that Floquet-dresses the electronic bands and gives rise to a finite transverse Hall response. The Floquet gaps can reach a substantial fraction of the driving frequency, $\Delta_F / \hbar\omega_c \sim 10^{-2}$ [see Fig. \ref{fig1}(b)], corresponding to gaps of order $\Delta_F \sim \CYNN{1-}10 \ \mathrm{meV}$ for near-IR frequencies. In this regime, the incoherent electron scattering time, with a characteristic time of $\sim 10 \ \mathrm{ps}$ \cite{Aeschlimann2021SurvivalScattering,fbe_orig,Liu2025}, is much longer than the timescale for coherent Floquet dynamics, $\hbar/\Delta_F \sim O(10 \ \mathrm{fs})$. This separation of timescales is enabled by the remarkably strong field amplitudes generated by the cavity. \CYNN{For} a typical near-IR frequency $\hbar\omega_c \sim 1 \ \mathrm{eV}$ \CY{on the order of typical semiconducting gap sizes}, and standard cavity volumes $\mathcal{V} \sim 10 \ \mathrm{\mu m}^3$, \CYNN{the system generates substantial cavity photon occupation numbers $n \sim 10^6$ and large electric field amplitudes reaching}
$\mathcal{E} \sim 0.1 \ \mathrm{MV/cm}$,  \CYNN{while still producing weak} heating power, $\sim 20 \ \mu\mathrm{W}$. 
\
{An additional key feature of the system in the strong-lasing regime is that the steady state is driven toward the ideal Floquet topological insulator (FTI) state, which exhibits maximal Hall conductivity. The reason is that the ideal FTI state corresponds to a distribution with no electronic population above the first photon resonance, precisely the same depletion of population inversion that leads to saturation of the cavity field [see Fig.~\ref{fig3}(d)].}

Conceptually, this work departs from the standard Floquet paradigm in which time-periodic driving is externally imposed and treated as an independent control parameter. Here, the drive amplitude and frequency are selected dynamically through feedback between light and matter and are stabilized by dissipation. As a result, \CY{drive-induced Berry flux} 
becomes a property of a self-organized nonequilibrium phase rather than a response to an externally tuned field. This perspective naturally connects ideas from Floquet engineering, cavity quantum electrodynamics, and laser physics, and highlights the constructive role of dissipation in stabilizing coherent nonequilibrium states.

From a practical standpoint, the cavity-mediated approach offers several advantages over conventional laser-driven schemes. \CY{The electromagnetic field is self-consistently generated by the driven material itself, so it is confined to the exact same volume set as the size of the material, enabling strong dressing of its electronic states. Additionally, the input power is set by the balance between photon gain and loss, rather than the large field intensity for the cavity field, enabling high efficiency}. 
\CY{Lastly, the} absence of external time-dependent fields simplifies device integration and allows the \IE{geometric} response to be probed using standard dc transport measurements. These features make electrically driven cavity platforms a promising route toward controllable nonequilibrium \IE{devices based on Floquet band geometry.} 

While our analysis has focused on a minimal model, the underlying mechanism is generic and can be extended in several directions. Different material platforms, including Dirac and Weyl systems, transition-metal dichalcogenides, and correlated narrow-band materials, provide natural settings where cavity-enhanced light--matter coupling can induce substantial Floquet effects. Multi-mode cavities and spatially structured electromagnetic fields offer additional opportunities to engineer more complex driven steady states, including spatially inhomogeneous or time-crystalline phases.

Several open questions remain. An important direction for future work is the role of strong electronic interactions, which may lead to new forms of correlated Floquet steady states beyond the single-particle picture considered here. Another natural extension concerns fluctuations and noise around the limit cycle, including spatiotemporal fluctuations of the cavity field, which may impact coherence, stability, and transport in experimentally relevant regimes. In this context, mode competition between different cavity or spatial modes may play an important role in selecting the steady state and could lead to richer dynamical phases.
Finally, the interplay between cavity-induced topology and edge or interface states, as well as the possibility of quantized responses in strongly gapped regimes, merit further investigation.

More broadly, our results suggest that self-organized nonequilibrium phases provide a fertile platform for realizing and controlling \CY{band geometry}.
By embedding quantum materials in structured electromagnetic environments and driving them electrically, it may be possible to access regimes of Floquet engineering that are difficult or impossible to reach with conventional external drives. \IE{We anticipate that this approach will stimulate further exploration of cavity controlled band geometry, Floquet topology, and nonequilibrium phases of matter.} 

\begin{acknowledgments}
We thank Mohammad Hafezi and \CYNN{Anatoli Polkovnikov} for insightful discussions. C.Y. gratefully acknowledges support from the Eddleman Quantum Institute postdoctoral fellowship and Moore Foundation postdoctoral fellowship. G.R. is grateful for support from the Simons Foundation, the Institute for Quantum Information and Matter, an NSF Physics Frontiers Center (PHY-2317110), and the AFOSR MURI program, under agreement number FA9550-22-1-0339. M.R. acknowledges the Brown Investigator Award, a program of the Brown Science Foundation, the University of Washington College of Arts and Sciences, and the Kenneth K. Young Memorial Professorship for support.
\end{acknowledgments}

\bibliography{Bibliography}

\section{Supplementary Information}

\subsection{Electron-Phonon Coupling}
The coupling matrix element takes the form
\begin{equation}
g_{\mathbf{q}}^a = D_a(\mathbf{q})\sqrt{\frac{\hbar |\mathbf{q}|}{2 \rho \varpi v_s}}, \qquad g_{\mathbf{q}}^o =
D_o \sqrt{\frac{\hbar}{2 \rho \varpi \omega_o}},
\end{equation}
where $D_a(\mathbf{q})$ and $D_o$ are the effective acoustic and optical phonon deformation potentials, respectively, $\rho$ the surface mass density, and $\varpi$ the unit cell area. Here, 
\begin{equation}
    D_a(\mathbf{q}) = D_a + D_{\mathrm{PE}} (\mathbf{q})
\end{equation}
where $D_a$ is the longitudinal acoustic (LA) phonon deformation potential, and $D_{\mathrm{PE}} (\mathbf{q})$ is the momentum-dependent matrix element for LA phonon interactions via piezoelectric coupling in the inversion symmetry broken TMD \cite{PhysRevB.87.235312,Wu2014,Zhu2014}. In our numerics, we specialize to the case of a constant matrix element $D_a(\mathbf{q}) = D_a$ for simplicity, to capture the general phenomenology of electron-phonon relaxation processes.

\subsection{Simulation Parameters}
In our simulations, we assumed a direct band gap of $\Delta = 1 \ \mathrm{eV}$ in the optically-active valley of the TMD and chose $v = 0.5 \times 10^{6} \ \mathrm{m/s}$ [see Eq.~(\ref{eq:eqhamil})]. We used a cavity resonant frequency of $\hbar\omega_c = 1.3 \ \mathrm{eV}$ and assumed an active material of total area $7 \ \mathrm{\mu m} \times 7 \ \mathrm{\mu} m$ and height $\lambda/2$, where $\lambda = 2\pi c/\omega_c$ is the resonant wavelength and $c/\sqrt{\varepsilon}$ is the speed of light in a dielectric medium with permittivity $\varepsilon = 5$, typical of bulk TMDs.

Our Floquet-Boltzmann equation simulations were carried out on an $133 \times 133$ 2D square momentum grid with $k_x ,k_y \in [-1.7 \times 10^{9}\ \mathrm{m}^{-1}, 1.7 \times 10^{9} \ \mathrm{m}^{-1}]$. We used an acoustic phonon deformation potential of $D_a = 2.5 \ \mathrm{eV}$, optical phonon deformation potential of $D_o = 25.5 \ \mathrm{eV}/\mathrm{Å}$, non-radiative recombination rate of $1/\Gamma^{\mathrm{rec}}=30 \ \mathrm{ps}$, lead tunneling rate of $1/\Gamma^{\text{lead}}=0.9 \ \mathrm{ps}$, acoustic phonon speed of $v_s= 5 \ \mathrm{km/s}$, optical phonon frequency of $\hbar \omega_o= 40 \ \mathrm{meV}$, sheet mass density of $\rho = 1.5 \times 10^{-6} \ \mathrm{kg/m^2}$, and unit cell area of $\varpi=5.24 \ \mathrm{Å}^2$. We note that the non-radiative recombination rate is characteristic of typical defect-assisted Auger processes in doped TMDs \cite{Wang2014}. Our simulations also assumed filtered leads with energy windows chosen to be $\varepsilon^T_{\mathrm{min}} = \varepsilon^B_{\mathrm{max}} = 0$, $\varepsilon^T_{\mathrm{max}} = -\varepsilon^B_{\mathrm{min}} = 800 \ \mathrm{meV}$. Finally, to compute the power balance and efficiency, we assumed a contact resistance of $R_{\mathrm{cont}}=7 \ \mathrm{G\Omega}$ and a leakage current resistance of $R_{\mathrm{leak}}=3000 \ \mathrm{\Omega}$.

\subsection{Boltzmann equation numerical solution and self-consistent steady state} \label{sec:numsol}

We determine the nonequilibrium steady state of the coupled electron–photon system by solving the Floquet–Boltzmann \cite{PhysRevA.92.062108,PhysRevA.91.033601,fbe_orig,fbe_adv,Parmee2020,PhysRevResearch.3.L012016} 
kinetic equations for the electronic occupations [Eq.~\eqref{eq:floquet_kinetic}] together with the cavity photon rate equation [Eq.~\eqref{eq:rate_eq}] in a fully self-consistent manner. The solution proceeds iteratively and treats the cavity field amplitude $A = \sqrt{n}$,
the Floquet spectrum, and the electronic distribution on equal footing.

For a given cavity photon number $n$\CY{, taken to be an effectively continuous variable in the large $n$ limit,}  
we first construct the effective time-periodic electronic Hamiltonian
$H_{\mathrm{el}}(\mathbf{k},t)$ and compute its Floquet quasienergies $\varepsilon_{\mathbf{k}\nu}$ and Floquet eigenstates $|\phi_{\mathbf{k}\nu}(t) \rangle
=
\sum_m |\phi_{\mathbf{k}\nu}^{m} \rangle e^{-im\omega_c t}$ truncated to a finite number of harmonics $-2 \leq m \leq 2$.
\CY{Here, we obtain the time-periodic Hamiltonian via minimal coupling $H_{\mathrm{el}}(\mathbf{k},t) = H_{\mathrm{el}}(\mathbf{k}+e\mathbf{\mathcal{A}}(t)/\hbar)$, where $\mathbf{\mathcal{A}}(t) = \mathcal{A}_0 (\cos(\Omega t),\sin (\Omega t))$ is the magnetic vector potential for the circularly polarized field, where $\mathcal{A}_0 = \mathcal{E}_0 |A|/\omega_c$.} Using the resulting Floquet eigenstates, we evaluate the transition rates entering the kinetic equation Eq.~\eqref{eq:floquet_kinetic}. Electron--phonon scattering rates are computed using Fermi’s golden rule expressed in the Floquet basis, which naturally accounts for both photon-conserving and Floquet--Umklapp processes.

Specifically, the transition rate from Floquet state $(\mathbf{k}',\nu')$ to $(\mathbf{k},\nu)$ induced by electron--phonon coupling can be approximated by Fermi's golden rule in the weak scattering limit, see Eq.~(\ref{eq:flcond}), and is given by 
\begin{equation}
\begin{split}
R^{\Delta m,\mathrm{ph}}_{\mathbf{k}\nu,\mathbf{k}'\nu'}
&=
\frac{2\pi}{\hbar}
\sum_{\mathbf{q},\zeta}
\big|
M^{(\Delta m)\zeta}_{\mathbf{k}\nu,\mathbf{k}'\nu'}(\mathbf{q})
\big|^2
\,
\times \\
&\times \delta\!\left(
\varepsilon_{\mathbf{k}\nu}
-
\varepsilon_{\mathbf{k}'\nu'}
-
\Delta m \hbar\omega_c
-
\hbar \omega_{\mathbf{q}}^\zeta
\right),
\end{split}
\label{eq:fgr_floquet}
\end{equation}
where $\omega_{\mathbf{q}}^\zeta$ is the phonon frequency of the optical ($\zeta = o$) and acoustic ($\zeta = a$) branches [see Eq.~\eqref{eq:PhononDispersion}] and $\Delta m$ labels the Floquet harmonic exchanged in the scattering process. The matrix elements
$M^{(\Delta m)\zeta}_{\mathbf{k}\nu,\mathbf{k}'\nu'}(\mathbf{q})$
are Floquet-{harmonic}-resolved electron--phonon coupling amplitudes, given by
\begin{equation}
M^{(\Delta m)\zeta}_{\mathbf{k}\nu,\mathbf{k}'\nu'}(\mathbf{q})
=
\sum_{m}
\langle
\phi^{m+\Delta m}_{\mathbf{k}\nu}
|
\hat H_{\mathrm{el\text{-}ph}}^\zeta(\mathbf{q})
|
\phi^{m}_{\mathbf{k}'\nu'}
\rangle ,
\end{equation}
where
$|\phi^{m}_{\mathbf{k}\nu}\rangle$ denotes the $m$-th Fourier component of the Floquet eigenstate. 

Recombination processes are incorporated as momentum-conserving interband relaxation terms, providing an additional channel for energy dissipation with the effective rate \cite{fbe_orig,transport} 
\begin{equation}
    R^{\Delta m,\text{rec}}_{\boldsymbol{k}\nu,\boldsymbol{k}\nu'} =  |S^{(\Delta m)}_{\boldsymbol{k}\nu\nu'} |^2 \Theta(\varepsilon_{\boldsymbol{k}\nu} - \varepsilon_{\boldsymbol{k}\nu'} - \Delta m \hbar\omega_c),
\end{equation}
where
\begin{equation} \label{eq:recme}
    S^{(\Delta m)}_{\mathbf{k}\nu\nu'} = \sum_{m} \langle \phi^{m+\Delta m}_{\mathbf{k}\nu} | \hat{G}(\mathbf{k}) | \phi^m_{\mathbf{k}\nu'}\rangle.
\end{equation}
Here, $\hat{G}(\mathbf{k})$ describes the electric dipole operator for radiative recombination, and $\Theta(\varepsilon)$ is the Heaviside step function. Although radiative recombination provides a natural recombination channel, it is typically much slower than other mechanisms, most notably non-radiative Auger processes \cite{Kioupakis2011,Wang2014}. To phenomenologically incorporate these faster processes without explicitly introducing an electron-electron collision integral, we replace the microscopic dipole matrix element by the effective form $\hat{G}(\mathbf{k}) = \sqrt{\Gamma^{\mathrm{rec}}} \sum_{\sigma\neq \sigma'} \hat c^{\dagger}_{\mathbf{k}\sigma'} \hat c_{\mathbf{k}\sigma}$, where $\sigma$ and $\sigma'$ label the conduction- and valence-band states of the undriven Hamiltonian 
[see Eq.~(\ref{eq:eqhamil})] \cite{fbe_orig,transport}. The rate ${\Gamma^{\mathrm{rec}}}$ is then chosen to approximately model the recombination timescale due to radiative or non-radiative (Auger) processes.

Lastly, tunneling to the leads is treated within the wide-band approximation, which gives Floquet-resolved tunneling rates $\Gamma^{\Delta m}_{\ell,\mathbf{k}\nu}$ that include photon-assisted sidebands, as defined in Eq.~\eqref{eq:lead_term}. The transition rate is obtained from Fermi's golden rule,
\begin{equation}
    \Gamma^{\Delta m}_{\ell,\mathbf{k}\nu} = \frac{2\pi}{\hbar} \sum_{\mathbf{p},\sigma} \left|Y^{(\Delta m)}_{\ell \mathbf{p}\mathbf{k}\sigma}\right|^2 \delta(\varepsilon_{\mathbf{k}\nu} + \Delta m \hbar\omega_c - \varepsilon_{\ell \mathbf{p}}),
\end{equation}
where 
\begin{equation}
    Y^{(\Delta m)}_{\ell \mathbf{p}\mathbf{k}\sigma} = t_{\ell \mathbf{p}\mathbf{k}\sigma} \langle \mathbf{k}\sigma | \phi^{\Delta m}_{\mathbf{k}\nu} \rangle .
\end{equation}
and $|\mathbf{k}\sigma \rangle$ is the Bloch state [see Eq.~(\ref{eq:eqhamil})]. We assume that the lead momentum can be decomposed as $\mathbf{p} = \mathbf{p}_{xy} + \mathrm{p}_z \hat{z}$, where $\mathbf{p}_{xy}$ is the in-plane component and $\mathrm{p}_z$ is the out-of-plane component, and that the tunneling conserves in-plane momentum, i.e.,
\begin{equation}
    t_{\ell \mathbf{p}\mathbf{k}\sigma} = J \delta_{\mathbf{p}_{xy}, \mathbf{k}}.
\end{equation}
The remaining sum over $\mathrm{p}_z$ then defines the lead density of states,
\begin{equation}
    \sum_{\mathrm{p}_z} \delta(\varepsilon_{\mathbf{k}\nu} + \Delta m \hbar\omega_c - \varepsilon_{\ell \mathbf{p}}) = \rho_{\ell}(\varepsilon_{\mathbf{k}\nu} + \Delta m \hbar\omega_c).
\end{equation}

In our calculations, we assume filtered leads \cite{fbe_orig,transport,PhysRevB.110.075428,PhysRevB.96.165443} 
with a constant density of states, with the top and bottom leads filtered to non-overlapping energy windows below and above $\varepsilon_{\mathbf{k}\nu}=0$, respectively, see Fig.~\ref{fig2}(a).
Such filtered leads can be realized using electrically and chemically biased narrow-bandwidth materials with a band gap larger than that of the TMD. Within this approximation,
\begin{equation}
    \rho_{\ell}(\varepsilon) =
    \begin{cases}
        \rho_0, & \varepsilon_{\mathrm{min}}^{\ell} < \varepsilon < \varepsilon_{\mathrm{max}}^{\ell}, \\
        0, & \mathrm{otherwise}.
    \end{cases}
\end{equation}
Here, $[\varepsilon_{\mathrm{min}}^{\ell}, \varepsilon_{\mathrm{max}}^{\ell}]$ denotes the filtered energy window, see Fig.~\ref{fig2}(a) for an illustration. Under these assumptions, the lead tunneling rate simplifies to
\begin{equation}
    \Gamma^s_{\ell,\mathbf{k}\nu} = \Gamma^{\text{lead}} \langle \phi^s_{\mathbf{k}\nu} | \phi^s_{\mathbf{k}\nu} \rangle,
\end{equation}
where
\begin{equation}
    \Gamma^{\text{lead}} = \frac{2\pi}{\hbar} |J|^2 \rho_0
\end{equation}
is a constant effective lead tunneling rate \cite{transport}.

For fixed {photon number} $n$, the steady-state electronic occupations $f_{\mathbf{k}\nu}$ are obtained by solving the stationary condition $\dot f_{\mathbf{k}\nu}=0$ \CY{in the Floquet-Boltzmann equation, see Eq.~(\ref{eq:floquet_kinetic})}.
This is done by solving the resulting system of nonlinear algebraic equations using a fixed-point iteration scheme. 
The resulting electronic distribution is generally nonthermal and explicitly depends on the cavity field amplitude through the Floquet spectrum.
Given the steady-state occupations, we evaluate the electronic contribution to the cavity photon dynamics by computing the net photon emission rate $\chi(n)$ from Eq.~\eqref{eq:photon_rate}. The steady-state photon number $n_\ast$ is then determined by solving the cavity rate equation, Eq.~\eqref{eq:rate_eq}.

\subsection{Details of the phenomenological model}
\label{sec:phenom_details}

\begin{figure}[t]
\includegraphics[width=0.6\columnwidth]{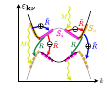}
\caption[]{\textbf{Phenomenological model.} 
Sketch of all phonon-mediated scattering processes (of intrinsic rates $R$ and $\tilde R$) and recombination processes (of intrinsic rate $M$) included in the phenomenological model.  }
\label{figs1}
\end{figure}

In this section, we provide details of the phenomenological model used in the main text. The model accounts for the occupations $F_i$ and $F_o$ of the inner and outer patches near the Floquet resonance and the cavity photon occupation $n$. The full set of phonon-mediated relaxation processes and recombination processes included in the model is sketched in Fig.~\ref{figs1}. We denote
\begin{equation}
    \zeta \equiv \left(\frac{\Delta_F}{\hbar\Omega}\right)^2 .
\end{equation}
Accounting for all photo-assisted electron--phonon processes, with each contribution weighted by the corresponding change in photon number, the gain function is given by
\begin{equation}
\begin{split}
    G(n)
    &= R\zeta\left[
    F_o(1-F_i)-F_i(1-F_o)+F_o^2-F_i^2
    \right]  \\
    &= R\zeta\,(F_o-F_i)(1+F_o+F_i).
\end{split}
\label{eq:photon_rate_supp}
\end{equation}

The corresponding rate equations for the patch occupations are
\begin{equation}
\begin{split}
    \dot{F}_o ={}& \Gamma(1-F_o)
    + R\zeta\left[
    F_i(1-F_o)-F_o(1-F_i)-F_o^2
    \right]  \\
    &- RF_iF_o - MF_o^2 ,
\end{split}
\label{eq:Fo_rate_supp}
\end{equation}
and
\begin{equation}
\begin{split}
    \dot{F}_i ={}& -\Gamma F_i
    + R\zeta\left[
    F_o(1-F_i)-F_i(1-F_o)-F_i^2
    \right]  \\
    &- RF_iF_o + M(1-F_i)^2 .
\end{split}
\label{eq:Fi_rate_supp}
\end{equation}
The rate $R$ describes phonon-mediated relaxation processes, $M$ denotes the phenomenological recombination rate, and the lead tunneling terms inject electrons into the outer patch $S_o$ with intrinsic rate $\Gamma$ and remove electrons from the inner patch $S_i$ with the same intrinsic rate.

We now derive the steady-state solution in the ideal limit of a perfect cavity with no photon loss. In this case, the photon steady-state condition is simply $G(n) =0$. For a nonzero cavity field, $\zeta>0$, Eq.~(\ref{eq:photon_rate_supp}) implies
\begin{equation}
    F_i = F_o \equiv F.
\end{equation}
Substituting this constraint into Eqs.~(\ref{eq:Fo_rate_supp}) and (\ref{eq:Fi_rate_supp}) gives
\begin{equation}
    0 = \Gamma(1-F) - \left[M+R(1+\zeta)\right]F^2 ,
\label{eq:Fo_equal_supp}
\end{equation}
and
\begin{equation}
    0 = -\Gamma F + M(1-F)^2 - R(1+\zeta)F^2 ,
\label{eq:Fi_equal_supp}
\end{equation}
and taking the difference of these two equations eliminates the phonon-mediated terms and gives
\begin{equation}
    \frac{\Gamma}{M} = (1-F)^2+F^2 .
\label{eq:F_condition_supp}
\end{equation}
Solving this equation yields
\begin{equation}
    F = \frac{1-\sqrt{2\Gamma/M-1}}{2}.
\label{eq:F_solution_supp}
\end{equation}
Here we have chosen the lower branch, since the upper branch would give an unphysical Floquet gap $\Delta_F$. This solution requires
\begin{equation}
    \frac{\Gamma}{M}>\frac{1}{2}.
\end{equation}
Furthermore, the condition $F \geq 0$ imposes the additional constraint $\Gamma/M\leq 1$.

The remaining steady-state equation determines the field amplitude. Using Eq.~(\ref{eq:Fi_equal_supp}), we find
\begin{equation}
    \zeta
    =
    \left(\frac{\Delta_F^*}{\hbar\Omega}\right)^2
    =
    \frac{(1-F)^2-(\Gamma/M)F-(R/M)F^2}{(R/M)F^2}.
\label{eq:eta_solution_supp}
\end{equation}
For simplicity, we define
\begin{equation}
    x = \sqrt{2\Gamma/M-1},
    \qquad
    y=\frac{1-x}{2}.
\end{equation}
Using Eq.~(\ref{eq:F_condition_supp}), the numerator in Eq.~(\ref{eq:eta_solution_supp}) can be written as
\begin{equation}
    (1-y)^2-\frac{\Gamma}{M}y
    =
    \frac{x(3+x^2)}{4}.
\end{equation}
Therefore
\begin{equation}
    \left(\frac{\Delta_F^*}{\hbar\Omega}\right)^2
    =
    \frac{R_c}{R}-1,
\label{eq:eta_Rc_supp}
\end{equation}
where
\begin{equation}
    \frac{R_c}{M}
    =
    \frac{x(3+x^2)}{(1-x)^2},
    \qquad
    x = \sqrt{2\Gamma/M-1}.
\label{eq:Rc_supp}
\end{equation}
A physical finite-field solution requires $(\Delta_F^*)^2>0$, and hence
\begin{equation}
    R<R_c.
\end{equation}
Thus, in the perfect-cavity limit, the phenomenological model admits a finite-field steady state with $F_i=F_o$ only when
\begin{equation}
    \frac{1}{2} < \frac{\Gamma}{M}\leq 1,
    \qquad
    R<R_c.
\end{equation}

\end{document}